\documentclass[12pt]{article}
\usepackage{graphics,cite,amssymb,epsfig,float,psfrag}
\RequirePackage{epsfig}
\RequirePackage{subfigure}
\newcommand{\beq}{\begin{equation}}
\newcommand{\eequ}{\end{equation}}
\newcommand{\eeq}{\end{equation}}
\def\bea{\begin{eqnarray}}
\def\eea{\end{eqnarray}}
\def\as{\relax\ifmmode\alpha_s\else{$\alpha_s${ }}\fi}
\def \pt{\relax\ifmmode{p_t}\else{$p_t${ }}\fi}
\def\nn{\nonumber}
\newcommand{\noi}{\noindent}

\def\mh{\hat{m}}

\def\as{{\alpha_s}}
\def\be{\begin{equation}}
\def\ee{\end{equation}}
\def\ba{\begin{eqnarray}}
\def\ea{\end{eqnarray}}

\newif\ifdtup

\def\eqal2#1{\,\vcenter{\openup1\jot
\caja   \ialign{\strut \hfil$\displaystyle{##}$&\hfil$
\displaystyle{{}##}$\hfil &$
\displaystyle{{}##}$\hfil\crcr#1\crcr}}\,}

\begin{document}

\title{
\begin{flushright}\normalsize
\vspace{0cm}
DESY 01-138 \\ 
hep-ph/0109232\\
September 2001
\vspace{2.cm}
\end{flushright}
\bf The Radiative rare decays $B\to K^{**}\gamma$ in the light-cone 
 QCD sum rule approach}

\author{A. S. Safir\footnote{e-mail~: safir@mail.desy.de} \\
        DESY, Deutsches Elektronen-Synchrotron, \\
        D-22603 Hamburg, Germany.}
\par \maketitle

%\end{center}

\begin{abstract}
We predict contributions of higher $K$-resonances to the radiative rare decays $b\to s\gamma$, in the framework of the QCD sum rules on the light cone, which combines the traditional QCD sum rule method with the description of final state mesons in terms of the light-cone wave functions of increasing twist. Our calculations are restricted to the leading twist-two operators for $K^*(892)$ and to the asymptotic wave function for the other $K^{**}$-mesons.

Using experimental data on the semileptonic $\tau\to K^{**}\nu_{\tau}$
decays, we extract the corresponding decay constants for vector and
axial-vector $K^{**}$-mesons. Based on our estimate of the transition
form factor $F_1^{K^*(892)}(0)=0.32 \pm 0.06$ and the $K_2^*(1430)$
decay constant $f_{K_2^*(1430)}=(160\pm 20)\ MeV$, we find good
agreement with experimental results. The two largest fractions of the
inclusive $b\to s\gamma$ branching ratio are found to be $(10.0\pm
4.0)\%$ for $B\to K^*(892) \gamma$ and $(5.0\pm 2.0)\%$ for $B\to
K_2^*(1430)\gamma$ decays. We also compare our results with the
existing theoretical predictions.

\end{abstract}

\newpage \pagestyle{plain}

\section{Introduction}
\hspace*{\parindent}
The flavour-changing neutral current (FCNC) $B$-decays, involving  the $b$-quark transition $b\to (s,d)+\gamma$ and $b\to (s,d)+l^+ l^-$ ($l= e,\ \mu,\tau,\nu$) provide a crucial testing grounds for the standard model at the quantum level, since such transitions are forbidden in the Born approximation. Hence, these rare $B$-decays are characterized by their high sensitivity to New physics.

In the standard model, the short distance contribution to rare B-decays is dominated by the top quark, and long distance contributions by form factors. Precise measurements of these transition will not only provide a good estimate of the top quark mass and the CKM matrix elements $V_{td},\ V_{ts},\ V_{tb}$, but also of the hadronic properties of $B$-mesons, namely form factors which in turn would provide a good knowledge of the corresponding dynamics and more hint for the non-perturbative regime of QCD.

The experimental searches for the FCNC B-decays have already provided first evidents. The recent experimental observations of the rare decay mode $B \to K^* \gamma$ have been determined by CLEO \cite{CLEO}, and more recently also by BABAR \cite{BABAR} and BELLE \cite{BELLE}

\begin{eqnarray}
Br(B^0 \to K^{*0} \gamma)=\cases{
(4.55\pm 0.70\pm 0.34)  \times 10^{-5}&\cite{CLEO} \cr 
(4.39\pm 0.41\pm 0.27)  \times 10^{-5}&\cite{BABAR} \cr 
(4.96\pm 0.67\pm 0.45)  \times 10^{-5}& \cite{BELLE} \cr}
\label{Brk*0}
\end{eqnarray}

and
\begin{eqnarray}
Br(B^+ \to K^{*+} \gamma)=\cases{
(3.76\pm 0.86\pm 0.28)  \times 10^{-5}&\cite{CLEO} \cr 
(3.89\pm 0.93\pm 0.41)  \times 10^{-5}&\cite{BELLE}\cr}
\label{Brk*+} 
\end{eqnarray}

and also of the inclusive rate \cite{CLEO2,ALEPH,BELLE2}:
\begin{eqnarray}
Br(B \to X_{s} \gamma)=(3.22\pm 0.40)\times 10^{-4}
\label{BrXs}
\end{eqnarray}

However, the first observation of the rare $B$-decay to the orbitally excited strange mesons has been reported only by CLEO \cite{CLEO}. With an integrated luminosity of $9.2\ fb^{-1}$ of $e^+e^-$ data corresponding to $9.7\times 10^6$ $B\bar{B}$ mesons pairs, they measure the $B\to K_2^*(1430) \gamma$ decay with a branching fraction of
\begin{eqnarray}
Br(B \to K_2^*(1430) \gamma)= (1.66 ^{+0.59}_{-0.53} \pm 0.13) \times 10^{-5}\label{Brk2*} 
\end{eqnarray}

These important experimental measurements provides a crucial challenge to the theory. Many theoretical approaches have been employed  to predict the exclusive $B\to K^*(892) \gamma$ decay rate (for a review see \cite{ali2} and references therein). On the other hand less attention has been devoted to rare radiative $B$-decays to excited strange mesons \cite{altomari,mannel,Veseli,Faustov}. Most of these theoretical approaches rely on non-relativistic quark models\cite{altomari,mannel}, HQET \cite{Veseli} and relativistic model\cite{Faustov}. However there is a large spread between different results, due to a different treatment of the long distance effects.

The aim of this paper is to estimate the contribution of higher $K$-resonances to the radiative rare decays $b \to s \gamma$. While the effective Hamiltonian approach allows to include short distances QCD corrections at scales $\mu^2 \simeq m_b^2$, estimates of the hadronic matrix elements of the relevant operators necessarily require some non-perturbative technique. We use here QCD sum rule \cite{Shifman}, which has been proved to be a powerful tool for such purposes.

In this paper, we use an elaborate version of the QCD sum rule, known as QCD sum rule on the light cone which has been developed originally for light quark systems in \cite{Balitsky, Braun}. This method was demonstrated to be very suitable for exclusive process. Moreover, it is technically much simpler than the standard QCD sum rule approach, where the basic correlator between the vacuum and a light meson state is expressed by universal quantities called light-cone wave functions for that particular light meson. These wave functions represent distributions of the light-cone momentum of the constituents, and can be classified by their twist defined as the difference between the canonical dimension and the Lorentz spin of the corresponding operator. Their asymptotic form is fixed by perturbative QCD, while the non-asymptotic effects at lower momentum scales can be estimated from QCD sum rules for two point correlators of appropriate currents \cite{Chernyak, Gorsky}.

The principal theoretical result of this paper is a sum rule for the electromagnetic penguin form factor appearing in the decays $B \to K^{**} \gamma$, where\ $K^{**}$ can be pseudoscalar, vector, scalar, axial-vector, tensor or pseudo-tensor. We derive the corresponding sum rule.

The paper is organized as follows. In section 2, we review the effective Hamiltonian for radiative $B$-decays, dominated by the electromagnetic penguins. In section 3, we define the form factors, which govern the exclusive rare $B$-decays to orbitally excited $K^{**}$-mesons. In section 4, we describe the method of the QCD sum rules on the light cone and derive our general sum rule for the form factors of the electromagnetic penguin operators entering in the decays $B \to V(A) \gamma$ and $B \to T(T_A) \gamma$ where $V (A)$ and $T(T_A)$ stand respectively for vector (axial-vector) and tensor (pseudo-tensor) $K^{**}$-meson. The wave functions and the decay constants which are needed to evaluate the sum rule are discussed in section 5, and the numerical estimates of the form factors and the predicted Branching fractions of $B \to K^{**} \gamma$ are presented in section 6. Finally, section 7 contains a summary and some concluding remarks.

\vskip 5mm %------------------------------------------------------

\section{\bf Effective Hamiltonian}
\hspace*{\parindent}
At the quark level, the rare semileptonic decay 
$b \to s  \gamma$ can be described in terms of the effective 
Hamiltonian obtained by integrating out the top quark and $W^\pm$ 
bosons: 
\begin{equation}
H_{eff} = -4 \frac{G_F}{\sqrt{2}}  V_{t s}^\ast  V_{tb}  
\sum_{i=1}^{8}  C_{i}(\mu)  O_i(\mu) \; . 
        \label{eq:he}
\end{equation}

\noi where $V_{ij}$ are the corresponding CKM matrix elements and $G_F$ is the Fermi coupling constant. The operator basis is defined as follows \cite{AlietGreub,Buras}:
\begin{eqnarray}
O_{1} &=& (\bar{s}_{L \alpha} \gamma_{\mu} b_{L \alpha})\ (\bar{c}_{L \beta} \gamma^{\mu} c_{L \beta})\nn \\
O_{2} &=& (\bar{s}_{L \alpha} \gamma_{\mu} b_{L \beta})\ (\bar{c}_{L \beta} \gamma^{\mu} c_{L \alpha})\nn \\
O_{3} &=& (\bar{s}_{L \alpha} \gamma_{\mu} b_{L \alpha}) {\sum_{q=u,d,s,c,b}} (\bar{q}_{L \beta} \gamma^{\mu} q_{L \beta})\nonumber\\
O_{4} &=& (\bar{s}_{L \alpha} \gamma_{\mu} b_{L \beta}) \sum_{q=u,d,s,c,b}(\bar{q}_{L \beta} \gamma^{\mu} q_{L \alpha})\nonumber\\
O_{5} &=& (\bar{s}_{L \alpha} \gamma_{\mu} b_{L \alpha}) \sum_{q=u,d,s,c,b} (\bar{q}_{R \beta} \gamma^{\mu} q_{R \beta})\nonumber\\
O_{6} &=& (\bar{s}_{L \alpha} \gamma_{\mu} b_{L \beta}) \sum_{q=u,d,s,c,b} (\bar{q}_{R \beta} \gamma^{\mu} q_{R \alpha})\nonumber\\
O_{7} &=& {e \over 16\ \pi^2}\bar{s}_{\alpha}\ \sigma_{\mu \nu}\ (m_{b} R+ m_{s} L)\ b_{\alpha}\ F^{\mu \nu}\nonumber\\
O_{8} &=& {g_{s} \over 16\ \pi^2}\bar{s}_{\alpha}\ T_{\alpha \beta}^{a}\ \sigma_{\mu \nu}(m_{b} R+ m_{s} L)\ b_{\alpha}\ G^{a \mu \nu}\label{Os}
\end{eqnarray}

\noi where $L/R \equiv {(1 \mp \gamma_5)}/2$, $\sigma_{\mu \nu}={i\over 2}[\gamma_{\mu},\gamma_{\nu}]$ and $q_{\mu}=(p_{B} -p)_{\mu}$ is the four-momenta transfer of the photon. $T^{a}$, $a=1....8$ are the generators of QCD, and $\alpha,\beta$ are $SU(3)$ color indices. Here $F^{\mu \nu}$ and $G^{a\mu \nu}$ denote the electromagnetic and chromomagnetic field strength tensor, respectively.
 The Wilson coefficients $C_i(\mu)$ are evaluated perturbatively at the $W$ scale and then evaluated down to the renormalization scale $\mu \sim m_b$ by the renormalisation group equation. Actually, they are known at the next-to-leading order \cite{munz}.

%%%%%%%%%%%%%%%%%%%%%%%%%%%%%%%%%%%%%%%%%%%%%
\section{\bf Matrix elements }
\hspace*{\parindent}
Our purpose in this section is to write the specific matrix elements for  $B \to K^{**}\gamma$ transitions, where $K^{**}$ denotes the higher $K$-resonances summarized in table \ref{K^**}. At the tree level only the magnetic moment operator $O_7$ contributes to the transition amplitudes. Thus, we are interested only in the following matrix elements

\begin{eqnarray}
\langle K^{**} | \Theta_{\mu\nu}   q^\nu | B \rangle, \ 
\langle K^{**} | \Theta_{\mu\nu}^5 q^\nu | B \rangle
\label{eq:def}
\end{eqnarray}

\noi where $\Theta_{\mu\nu}=\bar s\sigma_{\mu\nu}b$ and $ \Theta_{\mu\nu}^5=\bar s\sigma_{\mu\nu}\gamma_5 b$ are respectively the tensor and pseudo-tensor currents. For a generic radiative decay $B \to F \gamma$, where $F=P, V, S, A, T$ and $T_A$ stands for pseudoscalar, vector, scalar, axial-vector, tensor and pseudo-tensor respectively, one defines a transition form factor $F^{F}_{i}(q^2)$ as:
%\bar s\ \sigma_{\mu\nu}(\gamma_5) q^\nu  b

\begin{eqnarray}
\langle {P(S)(p)} | \Theta_{\mu\nu}^5(\Theta_{\mu\nu}) q^\nu  |
B(p_B)\rangle & = &0\label{eq:FPS0}\\
\langle {P(S)(p)} | \Theta_{\mu\nu}(\Theta_{\mu\nu}^5) q^\nu  |
B(p_B)\rangle & = & i {F^{P(S)}_1(q^2) \over m_B+m_{P(S)}}\left\{q^2 (p_B+p)_\mu-(m_B^2-m_{P(S)}^2) q_\mu \right\}\label{eq:FPS}\\
\langle {V(A)(p,\epsilon)} | \Theta_{\mu\nu}(\Theta_{\mu\nu}^5) q^\nu  |
B(p_B)\rangle & = & i\epsilon_{\mu\nu\rho\sigma}\  \epsilon^{*\nu}\
p_B^\rho\ p^\sigma \, 2 F^{V(A)}_1(q^2)\label{eq:FV}\\
\langle {V(A) (p,\epsilon)} | \Theta_{\mu\nu}^5(\Theta_{\mu\nu}) q^\nu  |
B(p_B)\rangle & = &  F^{V(A)}_2(q^2) \left\{ \epsilon^*_\mu
  (m_B^2-m_{{V(A)}}^2) - (\epsilon^* p_B) \,(p_B+p)_\mu \right\}\nonumber\\
 &+& F^{V(A)}_3(q^2)
(\epsilon^* p_B)  \left\{ q_\mu - \frac{q^2}{m_B^2-m_{{V(A)}}^2}\, (p_B+p)_\mu
\right\}
\end{eqnarray}

where $F^{V(A)}_1(0)=F^{V(A)}_2(0)$, and:

\begin{eqnarray}
\langle T(T_A)(p,\epsilon) | \Theta_{\mu\nu} (\Theta_{\mu\nu}^5) q^\nu  |
   B(p_B)\rangle & = & 
i\epsilon_{\mu\nu\rho\sigma}\ \epsilon^{*\nu \alpha}\ {{p_B}_\alpha \over m_B}
p_B^\rho\ p^\sigma \, 2 F^{T(T_A)}_1(q^2) \label{eq:FT} \\
\langle T(T_A)(p,\epsilon) | \Theta_{\mu\nu}^5 (\Theta_{\mu\nu})   q^\nu  |
   B(p_B)\rangle & = &  
F^{T(T_A)}_1(q^2) \left\{ \epsilon^*_{\alpha \beta}
{p_B^\alpha\ p_B^\beta \over m_B}\ (p_B +p)_\mu - \epsilon^*_{ \mu  \alpha}{p_B^\alpha \over m_B}( m_B^2- m_{{T(T_A)}}^2)\right\}\nonumber\\
&+& F^{T(T_A)}_2(q^2)\left\{ \epsilon^*_{\alpha \beta}
{p_B^\alpha\ p_B^\beta \over m_B}\ q_\mu - \epsilon^*_ { \mu \alpha}{p_B^\alpha \over m_B}q^2\right\}+F^{T(T_A)}_3(q^2) \nonumber\\
&\times& \epsilon^*_{\alpha \beta} {p_B^\alpha\ p_B^\beta \over m_B^2\ m_{T(T_A)}}\left\{ (m_B^2- m_{{T(T_A)}}^2)\ q_\mu - q^2 (p_B +p)_\mu \right\}
\end{eqnarray}

\noi where $m_{F}$, $\epsilon^{\mu}$ and $\epsilon^{\mu \nu}$ are respectively the generic mass of the $K^{**}$-meson, the polarization of the vector (axial-vector) and the tensor (pseudo-tensor) final meson. 

With the above definition, the exclusive decay widths are given by :
\begin{eqnarray}
\Gamma(B \to P(S)\gamma)& = &0\\
\Gamma(B \to V(A)\gamma)& = &  \zeta  m_{b}^5 |C_{7}(m_b)|^2 F^{V(A)}_1(0)^2 \left (1-{m_{V(A)}^2\over {m_B}^2}\right )^3 \left (1+ {m_{V(A)}^2\over {m_B}^2}\right )\ \\
\Gamma(B \to T(T_A)\gamma)& = &  {\zeta \over 8}\ m_{b}^5 |C_{7}(m_b)|^2 F^{T(T_A)}_1(0)^2\ {m_{B}^2 \over m_{T(T_A)}^2} \left (1-{m_{T(T_A)}^2 \over m_{B}^2}\right )^5 \nn \\
&&\left (1+ {m_{T(T_A)}^2 \over m_{B}^2}\right )
\end{eqnarray}

\noi with $\zeta = {\alpha\ G_{F}^2 \over 32 \pi^4}|V_{tb} V_{ts}^*|^2$. A good quantity to test the model dependence of the form factors for the exclusive decay is the ratio of the exclusive-to-inclusive radiative decay branching ratio \cite{Faustov}:
\begin{eqnarray}
R_{V(A)} &\equiv& {BR(B \to V(A)\ \gamma)\over BR(B \to X_s\ \gamma)} =   F^{V(A)}_1(0)^2 {\left (1-\mh_{V(A)}^2 \right )^3 \left (1+ \mh_{V(A)}^2 \right ) \over \left (1-m_{s}^2/ {m_b}^2\right )^3 \left (1+m_{s}^2/ {m_b}^2\right )}\label{eqRV}  \\
R_{T(T_A)} &\equiv& {BR(B \to T(T_A)\ \gamma)\over BR(B \to X_s\ \gamma)} ={ F^{T(T_A)}_1(0)^2\over 8 \mh_{T(T_A)}^2}{\left (1-\mh_{T(T_A)}^2 \right )^5 \left (1+ \mh_{T(T_A)}^2 \right ) \over \left (1-m_{s}^2/ {m_b}^2\right )^3 \left (1+m_{s}^2/ {m_b}^2\right )}\label{eqRT} 
\end{eqnarray}

\noi where $\mh_{F}= m_{F}/ m_{B}$. With this normalization, one eliminates the uncertainties from the CKM factor $V_{tb} V_{ts}^*$ and the short distance Wilson coefficient $C_{7}(m_b)$. Thus, we are left in eqs.(\ref{eqRV}) and (\ref{eqRT}) with unknown form factors $F^{F}_1(0)$, which we will derive in the next section.
\begin{table}
\begin{center}
\begin{tabular}{|cccc|}
%\hline & & & \\
\hline
Meson   &   $J^{P}$ &   $n^{2s+1}L_J$  &Mass (MeV) \\
\hline
$K$               &   $0^{-}$ & $1^1S_0$          & $497.67\pm 0.03$ \\
$K^{*}(892)$      &   $1^{-}$ & $1^3S_1$          & $896.1\pm 0.26$\\
$K_{1}(1270)$     &   $1^{+}$ & $1^1P_1/1^3P_1$   & $1273\pm7$\\
$K_{1}(1400)$     &   $1^{+}$ & $1^1P_1/1^3P_1$   & $1402\pm 7$\\
$K^{*}(1410)$     &   $1^{-}$ & $2^3S_1$          & $1414\pm 15$ \\
$K^{*}_{0}(1430)$ &   $0^{+}$ & $1^3P_0$          & $1412\pm 6$\\
$K^{*}_{2}(1430)$ &   $2^{+}$ & $1^3P_2$          & $1432\pm 1.3$ \\
$K(1460)$         &   $0^{-}$ & $2^1S_0$          & $\approx 1460$\\
$K_{2}(1580)$     &  $2^{-}$  & [$1^1D_2/1^3D_2$] & $\approx 1580$\\
$K_{1}(1630)$     &  $?^{?}$  &                   & $ 1629\pm 7$\\
$K_{1}(1650)$     &  $1^{+}$  & [$2^1P_1$]        & $ 1650\pm 50$ \\
$K^{*}(1680)$     &  $1^{-}$  & $1^3D_1$          & $1717\pm 27$\\
$K_{2}(1770)$     &  $2^{-}$  & $1^1D_2$          & $1773\pm 8$\\
$K^{*}_{3}(1780)$ &  $3^{-}$  & $1^3D_3$          & $1776\pm 7$\\
$K_{2}(1820)$     &  $2^{-}$  & $1^3D_2$          & $1816\pm 13$\\
$K(1830)$         &  $0^{-}$  & $3^1S_0$          & $\approx 1830$\\
$K^{*}_{0}(1950)$ &  $0^{+}$  & [$2^3P_0$]        & $1950\pm 10\pm 20$\\
$K^{*}_{2}(1980)$ &  $2^{+}$  & $2^3P_2$          & $1973\pm 8\pm 25$\\
$K^{*}_{4}(2045)$ &  $4^{+}$  & $1^3F_4$          & $2045\pm 9$\\
$K_{2}(2250)$     &  $2^{-}$  &                   & $2247\pm 17$\\
$K_{3}(2320)$     &  $3^{+}$  &                   & $2324\pm 24$\\
$K^{*}_{5}(2380)$ &  $5^{-}$  &                   & $2382\pm 14\pm 19$\\
$K_{4}(2500)$     &  $4^{-}$  &                   & $2490\pm 20$\\
$K(3100)$         &  $?^{?}$  &                   & $\approx 3100$\\
\hline
\end{tabular}
\end{center}
\caption{\it Spectrum of excited neutral $K$-meson taken from \cite{PDG}, except for the quantities in brackets. The quantum numbers $n$, $L$, $s$ and $J$ represent respectively the radial excitation, the orbital angular momentum, the sum of the two quark spins in the meson and the total meson spin. The parity is defined as $P=(-1)^{L+1}$. }\label{K^**}
\end{table}

%%%%%%%%%%%%%%%%%%%%%%%%%%%%%%%%%%%%%%%%%%%%%%%%%%%%%%%%%
\section{QCD sum rules on the light cone}
\hspace*{\parindent}
For definiteness, we show here and in the next section the derivation of the sum rule for  the transition form factor $F^{F}_1(0)$ as defined in eqs.(\ref{eq:FV}) and (\ref{eq:FT}), governing the radiative decays $B \to K^{**}\gamma$. 

The starting point of our sum rule is to consider the correlation function

\begin{eqnarray}
i \int {dx}\ e^{i qx} <K^{**}(p,\epsilon)|T \left \{\bar{\psi}(x) \sigma_{\mu \nu}(\sigma_{\mu \nu}\gamma_{5}) q^{\nu} b(x)\bar{b}(0)\ i \gamma_{5}\psi(0)\right\}|0> \label{eq01} 
\end{eqnarray}

\noi  which we will study in details for the vector (axial-vector) and tensor (pseudo-tensor) $K^{**}$-mesons. Hereafter we use $\psi$ as a generic notation for the field of the light quark. Our approach is very close to the calculation of the $B \to K(K^{*})$ form factor in \cite{Khodjamirian,AlietBraun}.

%%%%%%%%%%%%%%%%%%%%
\subsection{$B \to V(A)$ form factors}
\hspace*{\parindent}
To derive the sum rule, we start with the following matrix element of the time-ordered product of two currents between the vacuum state and the vector (axial-vector) $K^{**}$-meson at momentum $p$:

\begin{eqnarray}
i \int {dx}\ e^{i qx} <V(A)(p,\epsilon)|T \left \{\bar{\psi}(x) \sigma_{\mu \nu}(\sigma_{\mu \nu}\gamma_{5}) q^{\nu} b(x)\bar{b}(0)\ i \gamma_{5}\psi(0)\right\}|0> &=& i\ \varepsilon_{\mu \nu \rho \sigma}\epsilon^{*(\lambda)\nu} q^{\rho} p^{\sigma} \nn\\
&\times& T((p+q)^2)\label{eq:Vco} 
\end{eqnarray}

\noi at $q^2=0$  and at Euclidean $m_b^2-((p+q)^2)$ of order several $GeV^2$. The hadronic representation of (\ref{eq:Vco}) is obtained by inserting a complete set of states including the $B$-meson ground state, higher resonances and the non-resonant states with $B$-meson quantum numbers

\begin{eqnarray}
i\ \varepsilon_{\mu \nu \rho \sigma}\epsilon^{*(\lambda)\nu} q^{\rho} p^{\sigma} T((p+q)^2)&=& { \left < V(A)(p,\epsilon)|\bar{s}(x) \sigma_{\mu \nu}(\sigma_{\mu \nu}\gamma_{5}) q^{\nu} b(x)|B\right > \left< B |\bar{b}(0)\ i \gamma_{5}\psi(0)|0 \right >\over m_B^2-(p+q)^2 }\nn\\
&+& \sum_{h} { \left< V(A)(p,\epsilon)|\bar{s}(x) \sigma_{\mu \nu} 
    (\sigma_{\mu \nu}\gamma_{5}) q^{\nu} b(x)|h\right > 
               \left< h|\bar{b}(0)\ i \gamma_{5}\psi(0) |0\right> \over m_B^2-(p+q)^2}\nn\\
&&\label{eqhadrV}
\end{eqnarray}
Writing down the dispersion relation in $(p+q)^2$, we can separate the contribution of the $B$-meson as the pole contribution to the invariant function $ T((p+q)^2)$, i.e.

\begin{eqnarray}
 T((p+q)^2)={f_B \ m_B^2\over m_b}{2 F^{V(A)}_1(0)\over m_B^2-(p+q)^2}+.....
\label{eq:T}\end{eqnarray}

\noi where the dots stand for contributions of higher-mass resonances and the continuum. The $B$-meson decay constant is defined in the usual way,
\begin{eqnarray}
<0|\bar{\psi} \gamma_{\mu}\gamma_{5}b|B(p)>=i p_{\mu} f_B,
\end{eqnarray}

\noi $m_B$ and $m_b$ are the $B$-meson and the $b$-quark masses, respectively.

The possibility to calculate the correlator (\ref{eq:Vco}) in the region of large space-like momenta $(p+q)^2<0$ is based on the expansion of the $T$-product of the currents in (\ref{eq:Vco}) near the light-cone $x^2=0$. The leading contribution to the operator product expansion arises from the contraction of the $b$-quark operator in (\ref{eq:Vco}) to the free $b$-quark propagator $<0|b \bar{b}|0>$. The light quark operators are left uncontracted. Diagrammatically, this contribution is depicted in Fig.(1.a). The formal expression is easily obtained from (\ref{eq:Vco}):

\begin{eqnarray}
\int {dx}\ e^{i q x}\int{dk\over (2 \pi)^4}  e^{-i k x}\ {q^{\nu}\over m_b^2-k^2}\  <V(A)(p,\epsilon)|T \left \{ \bar{\psi}(x) \sigma_{\mu \nu}(\sigma_{\mu \nu}\gamma_{5})(m_b +\not{k}) \ i \gamma_{5}\psi(0)\right\}|0> \label{eq:for}
\end{eqnarray}

\noi where we made use of the following representation of the free propagator $\hat{S^0_b}(x)$:
\begin{eqnarray}
<0|T \left \{b(x)\bar{b}(0)\right\}|0> &=& i \hat{S^0_b}(x)=i \int{d^4p\over (2 \pi)^4} e^{-i p x} {m_b +\not{p}\over p^2-m_b^2}\nn\\
&=&-\int_0^{\infty}{d\alpha\over 16 \pi^2\alpha^2}\left (m_b +{i \not{x}\over 2\alpha}\right) e^{- m_b^2 \alpha+{x^2\over 4\alpha}}\label{eq:S0b}
\end{eqnarray}

In general, the expression (\ref{eq:for}) is expressed through matrix elements of non-local operators, sandwiched in between the vacuum and the meson state. These matrix elements define the light-cone meson wave functions. The first term in (\ref{eq:for}) is given by \cite{AlietBraun}:
\begin{eqnarray}
<0|\bar{\psi}(0) \sigma_{\mu \nu}(\sigma_{\mu \nu}\gamma_{5})\psi(x)|V(A)(p,\epsilon)>&=& i (\epsilon^{(\lambda)}_\mu\ p_\nu-\epsilon^{(\lambda)}_\nu\ p_\mu)\ f^\perp_{V(A)}\label{eq:1} \\ 
&&\times \int_0^1 {du}\ e^{-i u px} \phi_{\perp,V(A)}(u,\mu^2)\nn
\end{eqnarray}  

Likewise,
\begin{eqnarray}
<0|\bar{\psi}(0) \gamma_{\mu}(\gamma_{\mu}\gamma_{5})\psi(x)|V(A)(p,\epsilon)> &=&  f_{V(A)}m_{V(A)} \int_0^1 {du}\ e^{-i u px} \times \left[ p_{\mu} {(\epsilon^{(\lambda)}x)\over (p\ x)} \phi_{||,V(A)}(u,\mu^2)\right. \nn \\
%&\times& \left( p_{\mu} {(e^{(\lambda)}x)\over (px)} \phi_{||}(u,\mu^2)
&&\left. + \left(\epsilon^{(\lambda)}_{\mu}-p_{\mu} 
  {(\epsilon^{(\lambda)}x)\over 
  (p\ x)} \right) g^{(v)}_{\perp,V(A)}(u,\mu^2)\right]\label{eq:2}\\
%\int_0^1 {du}\ e^{-i u px} g^{V(A)}_{\perp}(u,\mu^2)\nn\\
<0|\bar{\psi}(0) \gamma_{\mu}\gamma_{5}(\gamma_{\mu})\psi(x)|V(A)(p,\epsilon)>&=& {-1\over 4} \epsilon_{\mu\nu\rho\sigma}  \epsilon^{(\lambda)\nu} p^\rho x^\sigma f_{V(A)}m_{V(A)}\nn\\
&& \times \int_0^1 {du}\ e^{-i u px} g^{(a)}_{\perp,V(A)}(u,\mu^2)\label{eq:3}
\end{eqnarray}  

The functions $\phi_{\perp,V(A)}(u,\mu^2)$ and $\phi_{||,V(A)}(u,\mu^2)$ give the leading-twist distributions in the fraction of total momentum carried by the quark in transversely and longitudinally polarized mesons respectively. The functions $g^{(v)}_{\perp,V(A)}(u,\mu^2)$ and $g^{(a)}_{\perp,V(A)}(u,\mu^2)$ are twist-3 wave functions. The normalization is chosen in such a way that for all all four distributions $f=\phi_{\perp,V(A)},\ \phi_{||,V(A)},\ g^{(v)}_{\perp,V(A)}$ and $g^{(a)}_{\perp,V(A)}$, we have:

\begin{eqnarray}
\int_0^1 {du}\ f(u)=1\label{eq:norma}
\end{eqnarray}

In the matrix elements of non-local operators on the l.h.s of (\ref{eq:1})-(\ref{eq:3}) the separations are assumed to be light-like, i.e. $x^2=0$. Regularization of ultraviolet divergences that arise in the process of the extraction of the leading $x^2 \to 0$ behavior produces a non-trivial scale dependence of the wave functions, which can be found by renormalization group methods. The scale in (\ref{eq:for}) is fixed by the actual light cone separation, $\mu^2 \sim x^{-2}\sim m_b^2-(p-q)^2$.

Putting eqs.(\ref{eq:for})-(\ref{eq:3}) together, we obtain:
\begin{eqnarray}
T((p+q)^2) &=& \int_0^1 {du} {1\over m_b^2+\bar{u}u m_{V(A)}^2-u(p+q)^2}\label{eq:4}\\
&&\times \left[m_b f^{\perp}_{V(A)} \phi_{\perp,V(A)}(u)+ u m_{V(A)} f_{V(A)} g_{\perp,V(A)}^{(v)}(u)+{1\over 4} m_{V(A)}f_{V(A)}g_{\perp,V(A)}^{(a)}(u)\right]\nn\\
&& + {1\over 4}\int_0^1 {du}{m_b^2-u^2\ m_{V(A)}^2\over( m_b^2+\bar{u}u m_{V(A)}^2-u(p+q)^2)^2}m_{V(A)}f_{V(A)}g_{\perp,V(A)}^{(a)}(u)\nn
\end{eqnarray}

\noi where $\bar{u}=1-u$. The expression in (\ref{eq:4}) has the form of a dispersion integral in $(p+q)^2$, which can be made explicit by introducing the squared mass of the intermediate state $s=m_b^2/u+\bar{u} m_{V(A)}^2$ as the integration variable, instead of the Feynman parameter $u$.The basic assumption of the QCD sum rule approach is that the contribution of the $B$-meson corresponds in this dispersion integral to the contribution of intermediate states with squared masses smaller than a certain threshold $s_0$ (duality interval). Making the Borel transformation $1/(s-(p+q)^2)\to exp(-s/t)$ and equating the result to the $B$-meson contribution in (\ref{eq:T}), we arrive at the sum rule

\begin{eqnarray}
{f_B \ m_B^2\over m_b}2 F^{V(A)}_1(0)e^{-(m_B^2-m_b^2)/t} &=& 
  \int_0^1 {du}\ {1\over u} exp \left[ 
  -{\bar{u}\over t}({m_b^2\over u}+m_{V(A)}^2) \right]
  \theta \left[s_0-{m_b^2\over u}-\bar{u} m_{V(A)}^2 \right]\nn\\
&\times& 
  \left[ m_b f^{\perp}_{V(A)} \phi_{\perp,V(A)}(u,\mu^2=t)+ u m_{V(A)}f_{V(A)}
  g_{\perp,V(A)}^{(v)}(u,\mu^2=t)\right. \nn\\
&+& 
  \left.  {m_b^2-u^2\ m_{V(A)}^2+ u t\over 4 u t}m_{V(A)}f_{V(A)}
  g_{\perp,V(A)}^{(a)}(u,\mu^2=t)\right]\label{eq:FSR}
\end{eqnarray}

\noi which should be satisfied for values of the Borel parameter $t$ of order several $GeV^2$. We see that Borel transformation removes arbitrary polynomials in $(p+q)^2$ and suppresses the contributions from excited and continuum states exponentially relative to the ground-state contribution.

%%%%%%%%%%%%%%%%%%%%%%%%%%%%%
\subsection{$B \to T(T_A)$ form factors}
\hspace*{\parindent}
The aim of this subsection is to calculate the transition form factor $F_1^{T(T_A)}$ as defined in (\ref{eq:FT}), governing the radiative decays $B\to T(T_A)\gamma$. We consider first the correlation function:

\begin{eqnarray}
F_{\mu}(q,p) = i \int{dx}\ e^{i qx} \langle T(T_A)(p,\epsilon)|T \left \{\bar{\psi}(x) \sigma_{\mu \nu}\gamma_{5}(\sigma_{\mu \nu}) q^{\nu} b(x)\bar{b}(0)\ i \gamma_{5}\psi(0) \right\}|0 \rangle \label{eq:Tco}
\end{eqnarray}

The hadronic representation of (\ref{eq:Tco}) is obtained in the same way as for (\ref{eq:Vco}). Finally, one gets 

\begin{eqnarray}
F_{\mu}(q,p)&=& \epsilon^*_{\alpha \beta}{p_B^\alpha\ p_B^\beta \over m_B}F(q^2,(p+q)^2) \ p_{\mu} + \epsilon^*_{\alpha \beta}{p_B^\alpha\ p_B^\beta \over m_B} H(q^2,(p+q)^2)\ q_{\mu}\nn\\
&+& \epsilon_ {\mu \alpha}^*{p_B^\alpha \over m_B}G(q^2,(p+q)^2)\label{eq2}
\end{eqnarray}

\noi where $q$ denotes the four-momentum transfer. From now on, we shall concentrate on the invariant amplitude $F$ which is physically more interesting than the amplitude $H$ and $G$. For $F$ one can write a general dispersion relation in the momentum squared $(p+q)^2$ of the $B$-meson:
\begin{eqnarray}
F(0,(p+q)^2)={f_B \ m_B^2\over m_b}\ {2 F^{T(T_A)}_1(0)\over m_B^2-(p+q)^2}+..
\label{eq2}\end{eqnarray}

\begin{figure}
\begin{center}
\epsfig{file=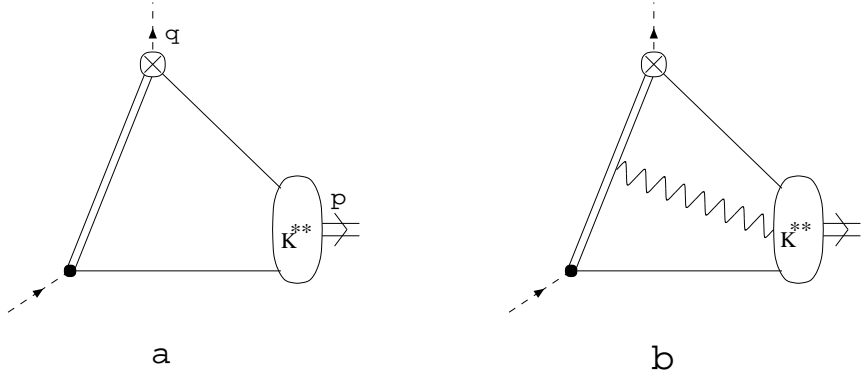,width=0.7\textwidth}
\caption{\it{The leading contribution {\bf a} and the gluon correction {\bf b} to the correlation function in (\ref{eq01}).}}
\label{fig1}
\end{center}
\end{figure}

From eqs.(\ref{eq:S0b}) and (\ref{eq:Tco}), one can obtain the leading contribution of Fig.(\ref{fig1}), and equals
\begin{eqnarray}
F_{\mu}(q,p) &=& \int_0^{\infty} {d\alpha\over 16 \pi^2\alpha^2}\ \int dx  e^{i q x-m_b^2\alpha+{x^2\over 4\alpha}}\label{eq:forT}\\ 
 &\times&  \left < T(T_A)(p,\epsilon)|T \left\{ \bar{\psi}(x) \sigma_{\mu \nu}\gamma_{5}(\sigma_{\mu \nu})q^{\nu} \left (m_b +{i \not{x}\over 2\alpha}\right) \ i \gamma_{5}\psi(0)\right\}|0\right> \nn
\end{eqnarray}

From now on, we will restrict ourself just to the helicity $\lambda=0$ for the tensor and axial-tensor mesons, due to the fact that this polarization brings mainly the leading twist-2 distribution. The other polarizations contribute in (\ref{eq:forT}) with more higher twist distributions, which will not be considered here. 

Let us consider the first term in (\ref{eq:forT}). The matrix element of the nonlocal operator is  given by \cite{Chernyak}:
\begin{eqnarray}
< T(T_A)(p,\epsilon)_{\lambda=0}| \bar{\psi}(x) \sigma_{\mu \nu}\gamma_{5}(\sigma_{\mu \nu})q^{\nu}\gamma_{5}\psi(0)|0>=-i p_{\mu} f_{T(T_A)}\int_0^1 {du}\ e^{i u q x}\phi_{T(T_A)}(u,\mu^2)\label{eq5}
\end{eqnarray}

\noi where $\phi_{T(T_A)}(u,\mu^2)$ is the tensor (pseudo-tensor) $K^{**}$-meson light-cone wave function of the leading twist-2, normalized as in (\ref{eq:norma}).
Substituting the matrix elements (\ref{eq5}) into (\ref{eq:forT}), and integrating over $x$ and the auxiliary parameter $\alpha$ we find the following expression for the coefficient $p_\mu$ in (\ref{eq2}) that is for the invariant amplitude $F$:

\begin{eqnarray}
F(q^2,(p+q)^2) = -f_{T(T_A)}m_b \int_0^1 {du} {\phi_{T(T_A)}(u,\mu^2)\over (p+q\ u)^2-m_b^2}\label{eq6}
\end{eqnarray}

After  using the Borel transformation of (\ref{eq6}) and equating to (\ref{eq2}),  the transition form factor $F^{T(T_A)}_1$ is obtained
\begin{eqnarray}
F^{T(T_A)}_1(0)={f_{T(T_A)}m_b^2 \over 2 f_B\ m_B^2}\int_\Delta^1 {du\over u}\ exp\left[{m_B^2 \over t}-{m_b^2 \over u t}\right]\phi_{T(T_A)}(u,\mu^2)\label{eq7}
\end{eqnarray}

\noi where $\Delta= (m_b^2/s_0)$ and $f_{T(T_A)}$ denotes the tensor (pseudo-tensor) decay constant.

In addition to the quark-antiquark wave functions considered above there are in principle also contributions from multi-particle wave functions. The most important corrections of this type are expected to arise from quark-gluon operators in the operator product expansion of (\ref{eq:Vco}) and (\ref{eq:Tco}). A typical diagram where the gluon is emitted from the heavy quark is shown in Fig.(1b). These contributions bring twist-3 contributions, which we do not take into account in this paper.

 Principal inputs in these sum rules are the vector (axial-vector) and the tensor (pseudo-tensor) meson wave functions and the meson decay constants, which encode non-trivial information about the dynamics at large distances, and which we are going to present now.

%%%%%%%%%%%%%%%%%%%%%%%%%%%%%%%%%%%%%%%%%%%%%%%%%%%%%%%%%
\section{Wave functions of the $K^{**}$-mesons}
\hspace*{\parindent}
Here, we collect the formulae for the light-cone wave functions of the $K^{**}$-mesons and specify the parameters. It is important to note that the asymptotic form of these functions and the scale dependence are given by perturbative QCD \cite{Chernyak}.
The twist-2 wave function $\phi_{\perp,K^*(892)}(u,\mu^2)$ is expressed as an expansion in Gegenbauer polynomials

\begin{eqnarray}
\phi_{\perp,K^*(892)}(u,\mu^2)= 6 u (1-u)\left [1 +a_1(\mu^2)\xi + a_2(\mu^2)\left(\xi^2-{1\over 5}\right)+a_3(\mu^2)\left ({7\over 3}\xi^3- \xi \right )+..\right ]\label{eq:phiK^*}
\end{eqnarray}

\noi where we have introduced the shorthand notation $\xi= 2u-1$. The non-perturbative effects are contained in the coefficients $a_{n}$. In the leading logarithmic accuracy, they are multiplicatively renormalizable and have the following scale dependence:

\begin{eqnarray}
a_n(\mu)= a_n(\mu_0) \left({\alpha_s(\mu)\over \alpha_s(\mu_0)} \right) ^{\gamma_n/b}\label{eq:an}
\end{eqnarray}

\noi where  $b=(11/3)N_c -(2/3)n_f$ is the LO coefficient of the QCD beta function, $N_c$ and $n_f$ being the number of colors and active flavours, respectively. The anomalous dimensions turn out to be \cite{AlietBraun}

\begin{eqnarray}
\gamma_n &=& C_F \left (1+ 4 \sum ^{n+1}_{j=2}{1/j} \right)
\end{eqnarray}

\noi where $C_F= (N_c^2-1)/2 N_c$. The non-asymptotic coefficient in (\ref{eq:phiK^*})  can be estimated from two-point sum rules \cite{Chernyak} for the moments $\int{u^n \ \phi_{\perp,K^*(892)}(u,\mu^2) du}$ at low $n$, or taken from experiment. The non-perturbative information encoded in the quark and gluon condensates is thereby transmuted into the long-distance properties of the wave function. For the numerical results, we use the following estimate at $\mu^2=1 \ GeV^2$ \cite{Chernyak,AlietBraun}

\begin{eqnarray}
a_{1}^{K^*(892)}=0.75,\ \ a_2^{K^*(892)}=-2,\ \ a_3^{K^*(892)}= 0.75 
\label{anvalues}
\end{eqnarray}

As mentioned above, the correct normalization point in the sum rules is the order of the typical Borel parameter, which for $B$-meson decays is of order $\mu^2\sim m_B^2-m_b^2\sim 6\ GeV^2$. Using $\Lambda^{(5)}_{\bar{MS}}= 0.225\ GeV$ with (\ref{eq:an}) and (\ref{anvalues}), one gets for the evoluted coefficients

\begin{eqnarray}
a_{1}^{K^*(892)}=0.57,\ \ a_2^{K^*(892)}=-1.35,\ \ a_3^{K^*(892)}= 0.46 \label{anvaluesat5}
\end{eqnarray} 

\begin{figure}
%\psfrag{c}{\hskip 0.3cm $s(GeV^2)$}
\psfrag{b}{\hskip -2. cm $\phi_{\perp}(u)$}
\psfrag{a}{\hskip 0.3cm $u$}
\begin{center}
\includegraphics[width=12cm,height=9cm]{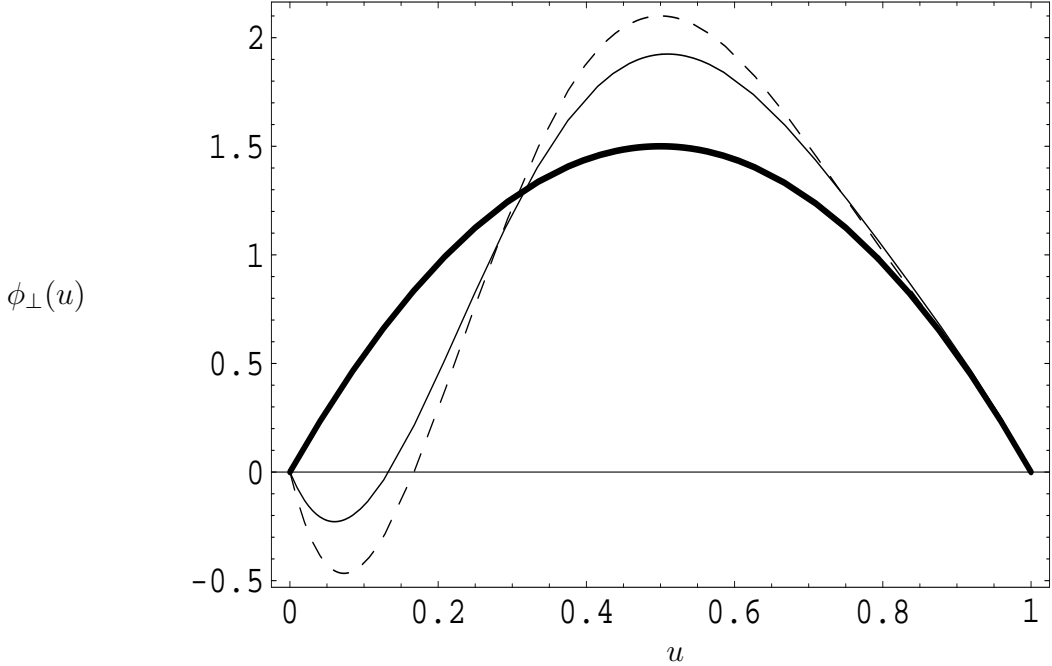}
\caption{\it{The leading twist-wave function $\phi_{\perp}(u)$ from the sum rule analysis in \cite{Chernyak,AlietBraun} for the $K^*$-meson. Solid and Dashed lines correspond respectively to the $K^*$-wave function at the scale $\mu^2=5.8$ and 1 $GeV^2$. The Thick one represents the asymptotic wave function $\phi_{\perp}(u)=6 u (1-u)$.}}
\label{figKwfct}
\end{center}
\end{figure}

The wave functions corresponding to the above parameterization are shown in Fig.(\ref{figKwfct}). As a general effect of the rescaling, the wave functions become somewhat wider and closer to the asymptotic expression $\phi_{\perp}(u)=6 u (1-u)$. In the case of the $K^*$-meson, the wave function becomes both wider and more symmetric, so the scaling violation affects the region $u<1/2$ only. Happily, the integration over $u$ in the sum rule in (\ref{eq:FSR}) is restricted to the interval of rather large momentum fractions, carried by the light quark involved in the electromagnetic penguin operator. For realistic values of parameters, one has for the $K^*$-meson $u>0.61-0.7$. According to the analysis in\cite{Chernyak}, the  $K^*$-wave function turns out to be in this region very close to the asymptotic expression.

The conformal expansion of the wave functions $g^{(v)}_{\perp,K^*(892)}(u,\mu^2)$ and $g^{(a)}_{\perp,K^*(892)}(u,\mu^2)$ is somewhat more involved and can be obtained using the approach\cite{Braun}. It should be noted that these wave functions contain contributions coming from both operators of twist-2 and twist-3. The twist-2 contributions to the transverse wave functions $g^{(v)}_{\perp,K^*(892)}$ and $g^{(v)}_{\perp,K^*(892)}$ receive a Wandzura-Wilczek type contribution \cite{Wandzura} that can be expressed in terms of the leading-twist longitudinal distribution function $\phi_{||,K^*(892)}$:

\begin{eqnarray}
g^{(v)}_{\perp,K^*(892)}(u)&=& {1\over2} \left[ \int_0^u {dw} {\phi_{||,K^*(892)}(w)\over 1-w}+ \int_u^1 {dw}{\phi_{||,K^*(892)}(w)\over w}\right]\nn\\
{d\over du} g^{(v)}_{\perp,K^*(892)}(u)&=& 2 \left[- \int_0^u {dw} {\phi_{||,K^*(892)}(w)\over 1-w}+ \int_u^1 {dw}{\phi_{||,K^*(892)}(w)\over w}\right]\label{WW}
\end{eqnarray}

In our evaluation we will not consider the twist-3 contributions to $g^{(v)}_{\perp,K^*(892)}(u,\mu^2)$ and $g^{(a)}_{\perp,K^*(892)}(u,\mu^2)$, which can be written in terms of three-particle antiquark-gluon-quark wave functions of transversely polarized vector meson \cite{Chernyak,Zhitnisky}. Following \cite{AlietBraun}, we write these functions

\begin{eqnarray}
g^{(v)}_{\perp,K^*(892)}(u)&=& {3\over 4} (1+\xi^2)\nn\\
g^{(a)}_{\perp,K^*(892)}(u)&=& {3\over 2} (1-\xi^2)\label{gva}
\end{eqnarray}

%\vspace{}

Concerning the other $K^{**}$-meson wave functions the situation is more complicated. It should be noted that the higher resonances are notoriously difficult to handle using QCD sum rules, because it is not possible to get rid in the sum rule of the contributions of the lower-lying states with the same quantum numbers. One possibility to solve this problem, as it is shown in Fig.(\ref{figKwfct}) is to use the asymptotic wave function for $K^{**}$-mesons. Since the light quark involved in the electromagnetic penguin operator for $B\to K^{**}$ transitions, carries a large momentum fractions $u>0.62-0.72$ in the sum rule (\ref{eq7}), then according to ref.\cite{Chernyak} the corresponding $K^{**}$-wave function in this region is very close to the asymptotic one. 

Thus, restricting our calculations for the $K^{**}$-wave functions to the asymptotic mode, we do not consider the higher twist contributions

\begin{eqnarray}
\phi_{\perp,V(A)}(u)&=& 6 \ u\ (1-u)\label{eq:phiT}\\
\phi_{T(T_A)}(u)&=& 6 \ u\ (1-u)\nn\\
g^{(v)}_{\perp,V(A)}(u)& = & 0\label{eq:gvforVT}\\
g^{(a)}_{\perp,V(A)}(u)& = & 0\nn
\end{eqnarray}

\noi Clearly, the asymptotic wave functions should be normalized to the characteristic scale of the process under consideration. Usually, this brings non-asymptotic effects into play which change the shape of the above wave functions, but preserve their normalization to unity. Unfortunately being of non-perturbative origin, these effects are difficult to evaluate.

%%%%%%%%%%%%%%%%%%%%%%%%%%%%%%%%%%%%%%%%%%%%%%%%%%%%%%%%%
\section{The decay constants}
\hspace*{\parindent}
The extraction of the $B$-meson decay constant $f_B$ from QCD sum rule has received a lot of attention in the past \cite{Reinders1,Aliev,Narison, Reinders2,Dominguez, Bagan, Neubert, Ruckl,Narison2}, using  for the two-point correlator of $\bar{b}\gamma_5 u$-currents.

The sum rule in \cite{AlietBraun} yields $f_B \sim 120-160\ MeV$, which is lower than the value preferred at present, typically $f_B \sim 180 \ MeV$ (see \cite{Dominguez,Ruckl} for a review). In fact this smaller value is an artifact of neglecting the radiative corrections, which are numerically large. Recently calculated perturbative three-loop QCD corrections are incorporated into the sum rule, leading to  $f_B =197\pm 23 \ MeV$ in ref.\cite{Jamin} and $\sim 206\pm 20 \ MeV$ in ref.\cite{Steinhauser}. Hence, we observe that the effect of the $O(\alpha_s)$ correction to the value of $f_B$ is sizeable. 

We note that what is actually calculated in (\ref{eq:FSR}) and (\ref{eq7}) are sum rules for the convoluted product of two amplitudes, $f_{B} F_1^F(0)$. From refs. \cite{AlietBraun} and \cite{Ruckl}, it is important to notice that $F_1^{K^*}(0)$ and $f_{B}$ increase either with higher operator contributions or with $O(\alpha_s)$ corrections. However, one can hope that going beyond the asymptotic distribution for $F_1^F(0)$ and $f_{B}$, the ratio of the corresponding sum rules will be stable.
Since we are discarding the $O(\alpha_s)$ radiative corrections in our sum rule and restricting our calculation to the leading twist-2 for $B\to K^{*}(892)\gamma$ decay and to the asymptotic distribution for $B\to K^{**}\gamma$ decays, we will take a reasonable value of $f_B \sim 160 \ MeV$ for $F_1^{K^*(892)}(0)$ and $f_B \sim 120 \ MeV$ for $F_1^{K^{**}}(0)$, as a guess of our estimate on the $B$-decay constant in our approach.

\begin{table}
        \begin{center}
        \begin{tabular}{|l|l|}
        \hline
        $m_{\tau}$                   & $1.777$ GeV \\   
%        $m_Z$                   & $91.1867$ GeV \\
%        $\sin^2 \theta_W $      & $0.2233$ \\
        $m_s$                   & $0.2$ GeV   \\
%        $m_d$                   & $0.01 $ GeV   \\
%        $m_c$                   & $1.4$ GeV \\
%        $m_{b pole}$            & $4.4 $ GeV \\
%       $m_t$                   & $173.8 \pm 5.0$ GeV     \\
%        $\mu$                & ${m_{b,pole}}$\\
        $\Lambda_{QCD}^{(5)}$   & $0.225$ GeV       \\
        $\alpha^{-1}$     & 137           \\
        $G_F$                   & $1.16637\ 10^{-5}\ GeV^{-2}$\\
%        $\alpha_s (m_Z) $       & $0.119 \pm 0.0058$ \\
%        ${\cal B}_{sl}$         & $(10.4 \pm 0.4)$ \%   \\
%        $\Gamma_{tot}(B)$       & $4.09 \cdot 10^{-13}$ GeV \\
%        $|V^\ast_{ts} V_{tb}|$ & 0.0385 \\
%        $|V_{ub}|/|V^\ast_{ts} V_{tb}|$ & 0.08\\
         $\tau_B $ & 1.61\ ps \\
         $t_{\tau}$ &0.291 \ ps \\
%        $\lambda_1$             & $-0.20$ GeV$^2$ \\
%        $\lambda_2$             & $+0.12$ GeV$^2$ \\
        \hline
        \end{tabular}
        \end{center}
\caption{\it Default values of the input parameters used in our numerical calculations.}
\label{parameters}
\end{table}

For the $K^{*}(892)$-meson decay constants, we adopt the same choice as in ref.\cite{AlietBraun}. Their numerical values determined from QCD sum rules are \cite{Shifman,Chernyak}:
\begin{eqnarray}
f_{K^*}^{\perp}=f_{K^*}=210 \ MeV\label{fk*}
\end{eqnarray}

In the case of $K^{**}$-mesons, the situation is more complicated since their decay constants, are not known, and we are not aware of any estimates of their numerical values. One should  extract them experimentally or rely on some models. Using the  semileptonic $\tau \to (P, V, S, A) \nu_{\tau}$ experimental data \cite{PDG}, it is possible to extract the corresponding meson decay constants. These quantities contribute to the transition rates for pseudoscalar, vector, scalar and axial vector emission, and we find from straightforward calculations \cite{Donogue}

\begin{eqnarray}
f_{P(S)} & =& \sqrt{Br^{exp}_{\tau \to P(S)\nu_{\tau}} {8 \pi \ m_{\tau} 
  \over t_{\tau}\ G_{F}^2 V_{us}^2 (m_{\tau}^2 - m_{P(S)}^2)^2}},\ \\
f_{V(A)} & =&\left( 1- {m_{V(A)}^2 \over m_{\tau}^2} \right)^{-1} \sqrt{Br^{exp}_{\tau \to V(A)\nu_{\tau}}{8 \pi \ m_{V(A)}^2 \over t_{\tau}\ G_{F}^2 V_{us}^2 m_{\tau}^3} \left (1+2 {m_{V(A)}^2 \over m_{\tau}^2} \right)^{-1}},\
\end{eqnarray}

\noi where $t_{\tau}$ is the $\tau$-lifetime, the quantities $f_{P(S)}$ and $f_{V(A)}$ are respectively the pseudoscalar (scalar) and the vector (axial-vector) meson decay constants. Using the central values of $Br^{exp}_{\tau \to (P,V,S,A)\nu_{\tau}}$ listed in \cite{PDG}, we obtain in table \ref{tab:p1} the corresponding decay constants. 

Concerning the tensor and the pseudo-tensor meson decay constants, it is difficult to extract them from the $\tau$ decay data, due to the impossibility to define them from the vacuum-to-tensor (pseudo-tensor) matrix element. A somewhat different approach can be used to obtain these missing decay constants, either using the SU(3) symmetry breaking or from other processes. 

The $K_{2}^*(1430)$ decay constant can be estimated from the experimental data on ($B\to K_{2}^*(1430)\gamma$) reported recently by CLEO \cite{CLEO}. Assuming a stability of our sum rule around $t=20 \ GeV^2$ and $t=25 \ GeV^2$, we plot in Figs.(\ref{fig3}) and (\ref{fig4}) the constraint on $f_{K_{2}^*(1430)}$. The prediction of the $K_{2}(1580)$ decay constant is somehow unknown and neither SU(3) symmetry breaking nor other processes are helpful. For that we assume a stability of our sum rule around $t=20\ GeV^2$ and $t=25 \ GeV^2$ , and we plot in Figs.(\ref{F10ofK21580}) and (\ref{25F10ofK21580}) the corresponding fractions of the inclusive $b\to s\gamma$ branching ratio (\ref{eqRT}) as function of the decay constant.

\begin{table}
\addtolength{\arraycolsep}{3pt}
\renewcommand{\arraystretch}{1.4}
$$
\begin{array}{|l|llllll|}
\hline
%&K(493)&K^*(892) 
&K^*_{1}( 1270)  & K_{1}(1400)&K^*(1410)&K^*(1430)&K_{1}(1650)&K^*(1680)\\ 
\hline
%&0^-&1^-
J^P   &1^+ & 1^+&1^-& 0^+&1^+&1^-\\
%&0.120& 0.154
f_{i}(MeV)  & 122 & 91
& 86 & 79& 86& 86 \\
\hline
\end{array}
$$
\caption[]{\it Central values of the pseudoscalar, vector, scalar and axial vector $K^{**}$-meson decay constants}\label{tab:p1}
 \end{table}

%%%%%%%%%%%%%%%%%%%%%%%%%%%%%%%%%%%%%%%%%%%%%%%%%%%%%%%%%
\section{Numerical results}
\hspace*{\parindent}
Having the necessary formulae at hand we next explain our choice of the relevant parameters. Let us first consider the parameters which characterize the $B$-meson channel that is $m_B= 5.27\ GeV$, $f_B$, $m_b$ and $s_0$. The value of the pole $b$-quark mass $m_b$ and the continuum threshold $s_0$, are varied in the limits:
\begin{eqnarray}
m_b&=& 4.6-4.8 \ GeV,\nn \\
s_0&=& 33-35\ GeV^2. \label{eq:mb,s0}
\end{eqnarray}

Before giving numerical predictions on the form factors $F_1^F(0)$ we must first determine the range of values for the Borel parameter $t$ for which the sum rules (\ref{eq:FSR}) and (\ref{eq7}) can be expected to yield reliable results. Usually, the lower limit of this range is determined by the requirements on the terms proportional to $t^{-n},\ n\geq 1$ to remain subdominant. In (\ref{eq:FSR}) this concerns in particular the twist-3 wave functions, which increase rapidly at small $t$. The upper limit of the allowed  interval in $t$ is determined by demanding the higher resonance and continuum contribution  not to grow too large. We  have checked numerically for $5 \leq t \leq 10\ GeV^2$ that the resulting value of $F_1^{K^*(892)}(0)$ is practically independent of the Borel parameter, as it is shown in Figs.(\ref{F10ofK892}) and (\ref{33F10ofK892}). 
%In addition we have studied the uncertainty on the parameters (\ref{anvaluesat5})

For other form factors involving $B\to V(A)\gamma$ transitions, we have used the asymptotic wave function (\ref{eq:phiT}), which  induces good stability for $F_1^{V(A)}(0)$ form factors, in the same Borel range.

The stability plots for the form factors as functions of the Borel parameter are shown in Figs.(\ref{F10ofK892}-\ref{33F10ofKstar1680}). The stability is in all cases good, and the variations of the continuum threshold within the limit (\ref{eq:mb,s0}) induce uncertainties in the values of form factors within less than $22\%$ for $F^{V(A)}_1$. Using the region in the Borel parameter $ 5\ GeV^2\ < t < 10\ GeV^2 $, we extract the values of the vector and the axial-vector form factors

\begin{eqnarray}
F_1^{K^*(892)}(0)&=&0.32 \pm 0.06 \nn\\
F_1^{K_1(1270)}(0)&=&0.14\pm 0.03 \nn\\
F_1^{K_1(1400)}(0)&=&0.098\pm 0.02\nn\\
F_1^{K^*(1410)}(0)&=&0.094\pm0.02\nn\\
F_1^{K_1(1650)}(0)&=&0.091\pm0.02\nn\\
F_1^{K^*(1680)}(0)&=&0.091\pm0.02\label{F1V(A)}
\end{eqnarray}

\begin{figure}
%\psfrag{c}{\hskip 0.3cm $s(GeV^2)$}
\psfrag{b}{\hskip -3. cm $R_{R_ {K^*_2(1430)}} [\%]$}
\psfrag{a}{\hskip 0.3cm $f_{K^*_2(1430)}(GeV)$}
\begin{center}
\includegraphics[width=12cm,height=9cm]{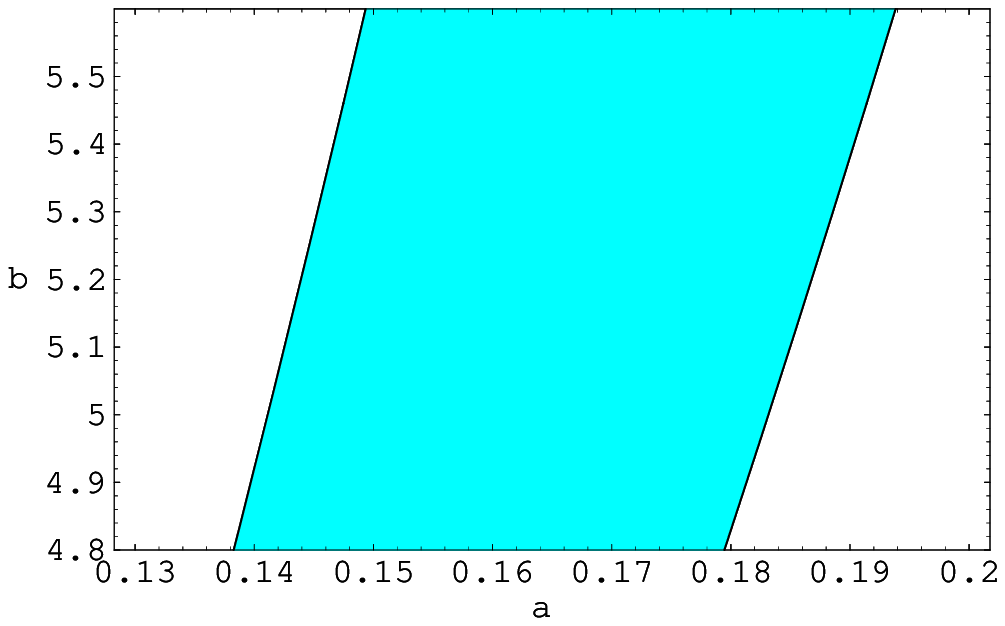}
\caption{\it{Experimental constraint on $f_{K^*_2(1430)}$ at $t=20 GeV^2$ and $s_0=35 GeV^2$.}}
\label{fig3}
\end{center}
%\end{figure}
\vspace{1.5cm}
%\begin{figure}
%\psfrag{c}{\hskip 0.3cm $s(GeV^2)$}
\psfrag{b}{\hskip -3. cm $R_{R_ {K^*_2(1430)}}[\%]$}
\psfrag{a}{\hskip 0.3cm $f_{K^*_2(1430)}(GeV)$}
\begin{center}
\includegraphics[width=12cm,height=9cm]{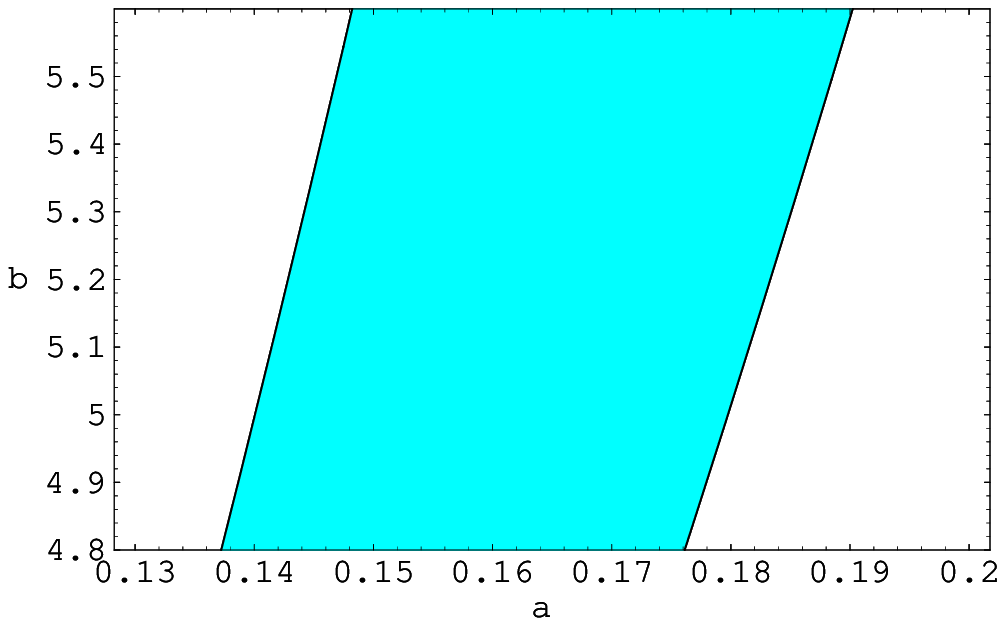}
\caption{\it{Experimental constraint on $f_{ K_2(1430)}$ at $t=25 GeV^2$ and $s_0=35 GeV^2$.}}
\label{fig4}
\end{center}
\end{figure}

Using the sum rule (\ref{eq7}), and the experimental data on $B \to K_2^*(1430)\gamma $ \cite{CLEO}, we show in Figs.(\ref{fig3}) and (\ref{fig4}) the experimental constraint on the $K_2^*(1430)$-decay constant. The shaded area represents the uncertainty on the $b$-mass. Assuming a stability of our sum rule  (\ref{eq7}) at $t= 20 \ GeV^2$ and $25 \ GeV^2$, we obtain 
\begin{eqnarray}
f_{K_2^*(1430)_{\lambda=0}}= (160 \pm 20) \ MeV \label{fk2star}
\end{eqnarray}

\noi Using the central value of (\ref{fk2star}), it turns out that the tensor form factor $F_1^{K^*_2(1430)}(0)$, as it is shown in Figs.(\ref{35F10ofK21430}) and (\ref{33F10ofK21430}), leads to
\begin{eqnarray}
F_1^{K^*_2(1430)_{\lambda=0}}(0)= 0.19 \pm 0.04 \label{F1T}
\end{eqnarray}

\begin{figure}
%\psfrag{c}{\hskip 0.3cm $s(GeV^2)$}
\psfrag{b}{\hskip -3.0cm $R_{K_2(1580)} [\%]$}
\psfrag{a}{\hskip -0.9cm $f_{K_2(1580)}(GeV)$}
\begin{center}
\includegraphics[width=12cm,height=9cm]{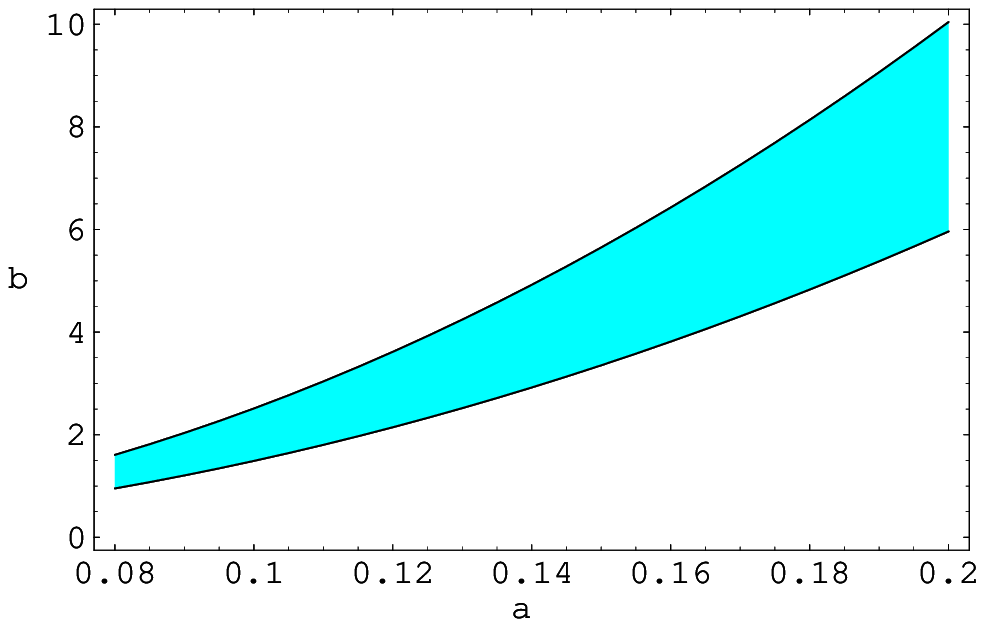}
\caption{\it Stability plots for  $f_{ K_2(1580)}$ at 
             $t=20 GeV^2$ and $s_0=35 GeV^2$.} \label{F10ofK21580}
\end{center}
%\hspace*{\parindent}
\vspace{1.5cm}
%\end{figure}
%\begin{figure}
%\psfrag{c}{\hskip 0.3cm $s(GeV^2)$}
\psfrag{b}{\hskip -3. cm $R_{K_2(1580)} [\%]$}
\psfrag{a}{\hskip -0.9cm $f_{K_2(1580)}(GeV)$}
\begin{center}
\includegraphics[width=12cm,height=9cm]{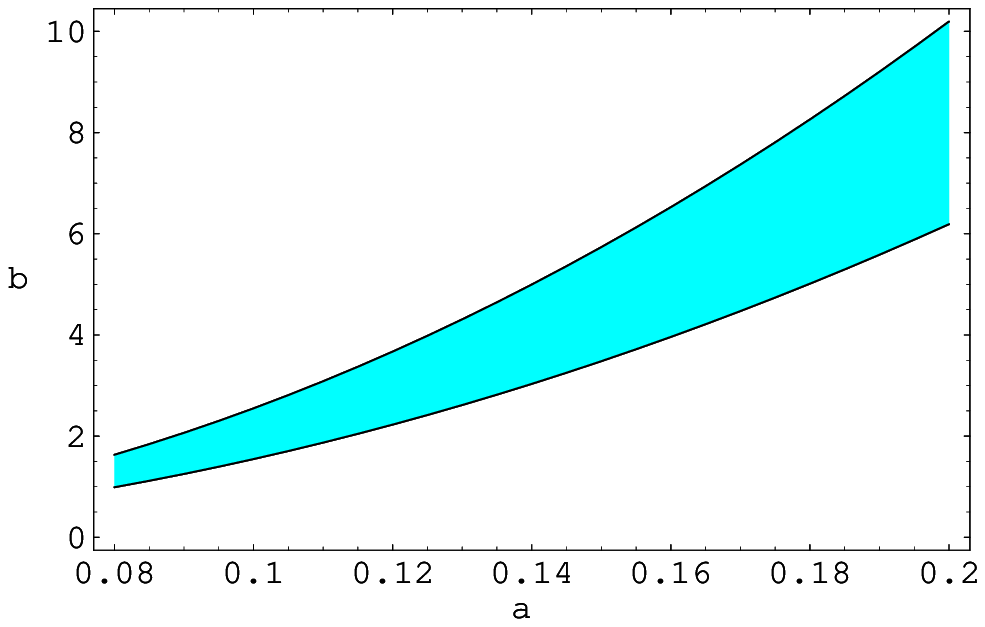}
\caption{\it Stability plots for  $f_{ K_2(1580)}$ at $t=25 GeV^2$ and 
          $s_0=35 GeV^2$.}
\label{25F10ofK21580}
\end{center}
\end{figure}

\begin{figure}
%\psfrag{c}{\hskip 0.3cm $s(GeV^2)$}
\psfrag{b}{\hskip -2. cm $F_1^{K^*(892)}(0)$}
\psfrag{a}{\hskip 0.3cm $t(GeV^2)$}
\begin{center}
\includegraphics[width=12cm,height=9cm]{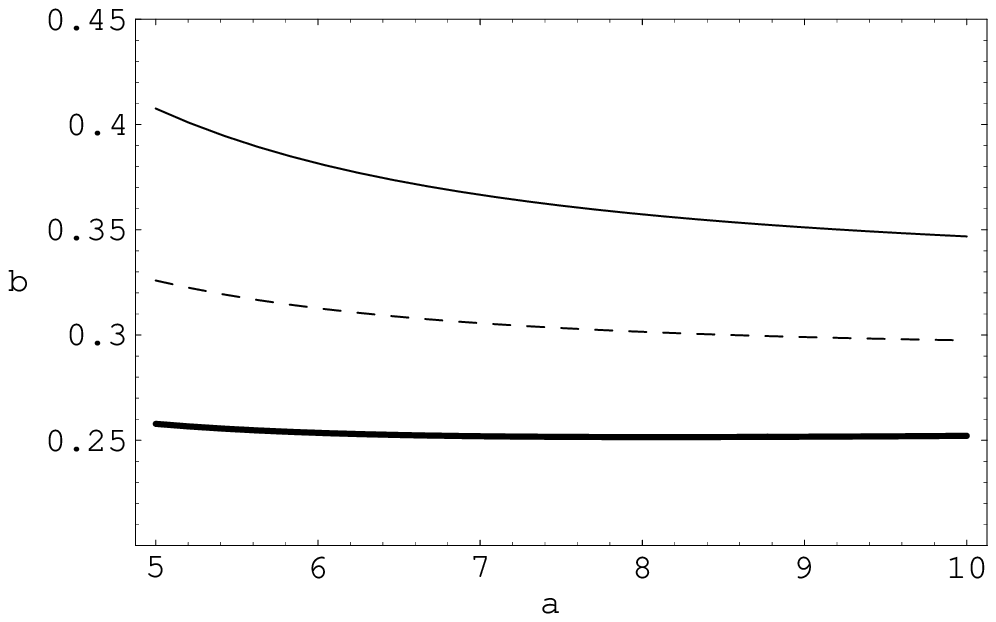}
\caption{\it{Stability plots for the sum rule in (\ref{eq:FSR}) of the Borel parameters for $B \to K^*(892)\gamma$ at $s_0=35\ GeV^2$. Solid, Dashed and Thick lines correspond respectively to $m_b$=4.6 GeV, 4.7 GeV and 4.8 GeV.}}
\label{F10ofK892}
\end{center}
\vspace{1.5cm}
%\end{figure}
%\begin{figure}
%\psfrag{c}{\hskip 0.3cm $s(GeV^2)$}
\psfrag{b}{\hskip -2. cm $F_1^{K^*(892)}(0)$}
\psfrag{a}{\hskip 0.3cm $t(GeV^2)$}
\begin{center}
\includegraphics[width=12cm,height=9cm]{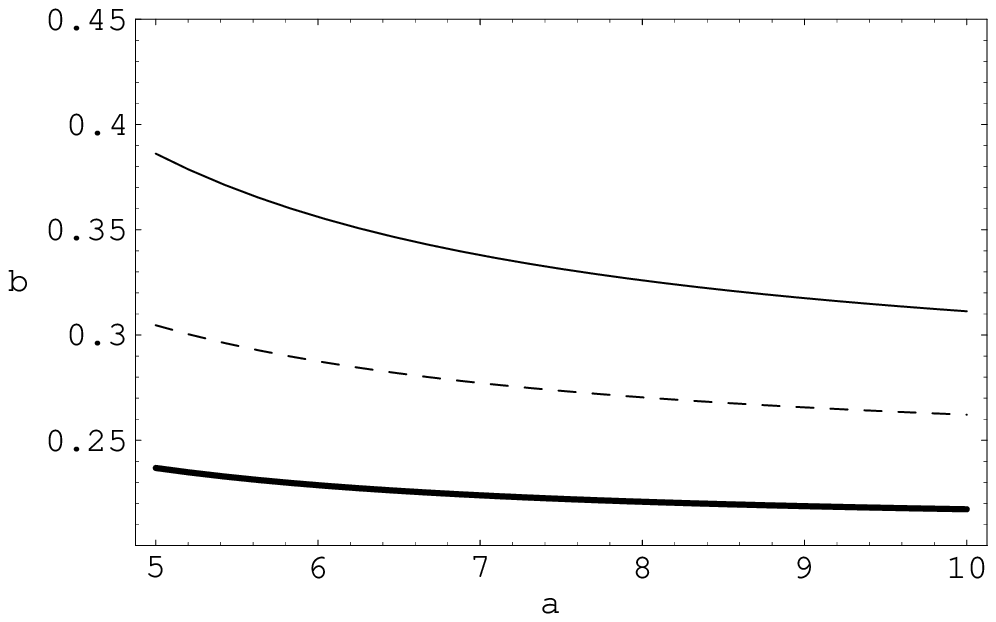}
\caption{\it{Stability plots for the sum rule in (\ref{eq:FSR}) of the Borel parameters for $B \to K^*(892)\gamma$ at $s_0=33\ GeV^2$. Solid, Dashed and Thick lines correspond respectively to $m_b$=4.6 GeV, 4.7 GeV and 4.8 GeV.}}
\label{33F10ofK892}
\end{center}
\end{figure}

\begin{figure}
%\psfrag{c}{\hskip 0.3cm $s(GeV^2)$}
\psfrag{b}{\hskip -2. cm $F_1^{K_1(1270)}(0)$}
\psfrag{a}{\hskip 0.3cm $t(GeV^2)$}
\begin{center}
\includegraphics[width=12cm,height=9cm]{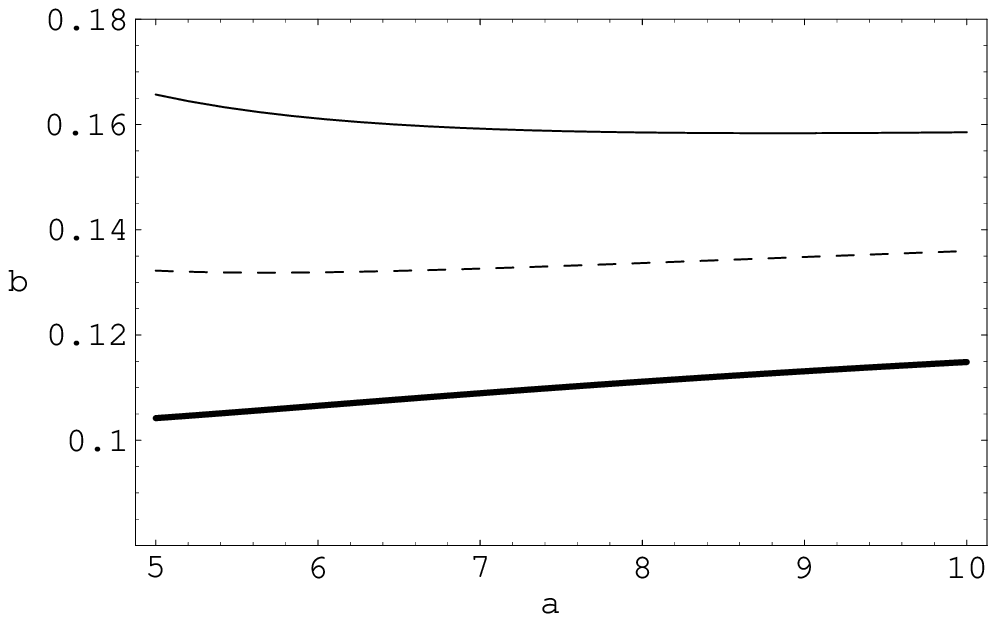}
\caption{\it{Stability plots for the sum rule in (\ref{eq:FSR}) of the Borel parameters for $B \to K_1(1270)\gamma$ at $s_0=35\ GeV^2$. Solid, Dashed and Thick lines correspond respectively to $m_b$=4.6 GeV, 4.7 GeV and 4.8 GeV.}}
\label{F10ofK11270}
\end{center}
%\end{figure}
\vspace{1.5cm}
%\begin{figure}
%\psfrag{c}{\hskip 0.3cm $s(GeV^2)$}
\psfrag{b}{\hskip -2. cm $F_1^{K_1(1270)}(0)$}
\psfrag{a}{\hskip 0.3cm $t(GeV^2)$}
\begin{center}
\includegraphics[width=12cm,height=9cm]{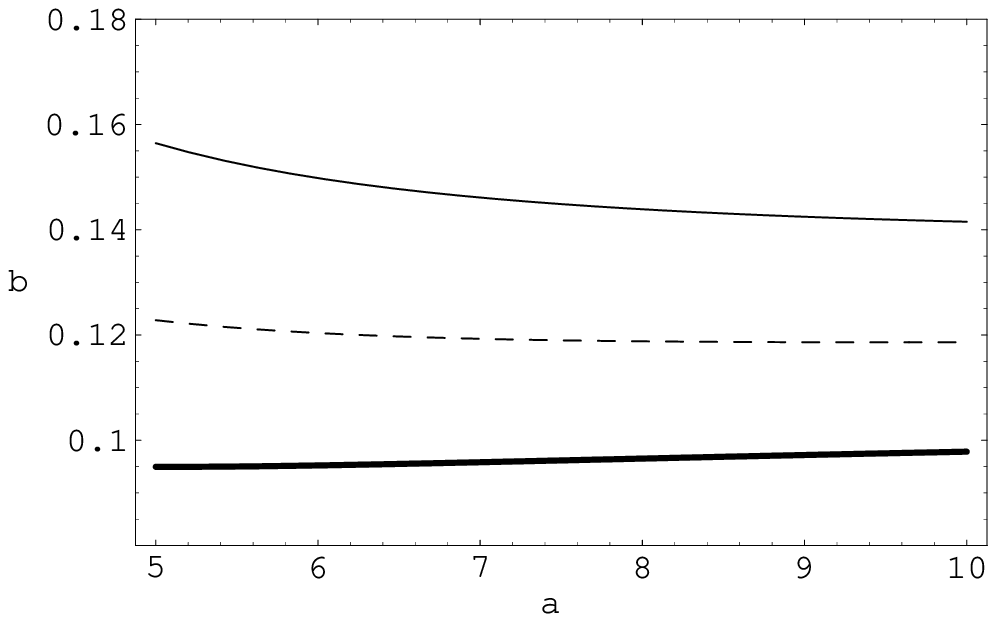}
\caption{\it{Stability plots for the sum rule in (\ref{eq:FSR}) of the Borel parameters for $B \to K_1(1270)\gamma$ at $s_0=33\ GeV^2$. Solid, Dashed and Thick lines correspond respectively to $m_b$=4.6 GeV, 4.7 GeV and 4.8 GeV.}}
\label{33F10ofK11270}
\end{center}
\end{figure}

\begin{figure}
%\psfrag{c}{\hskip 0.3cm $s(GeV^2)$}
\psfrag{b}{\hskip -2. cm $F_1^{K_1(1400)}(0)$}
\psfrag{a}{\hskip 0.3cm $t(GeV^2)$}
\begin{center}
\includegraphics[width=12cm,height=9cm]{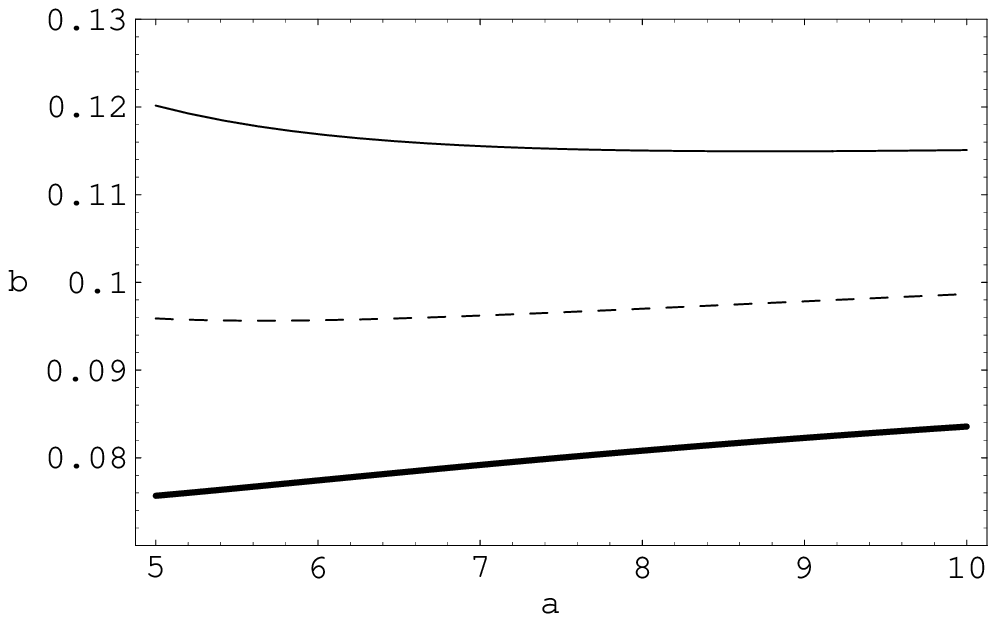}
\caption{\it{Stability plots for the sum rule in (\ref{eq:FSR}) of the Borel parameters for $B \to K_1(1400)\gamma$ at $s_0=35\ GeV^2$. Solid, Dashed and Thick lines correspond respectively to $m_b$=4.6 GeV, 4.7 GeV and 4.8 GeV.}}
\label{F10ofK11400}
\end{center}
%\end{figure}
\vspace{1.5cm}
%\begin{figure}
%\psfrag{c}{\hskip 0.3cm $s(GeV^2)$}
\psfrag{b}{\hskip -2. cm $F_1^{K_1(1400)}(0)$}
\psfrag{a}{\hskip 0.3cm $t(GeV^2)$}
\begin{center}
\includegraphics[width=12cm,height=9cm]{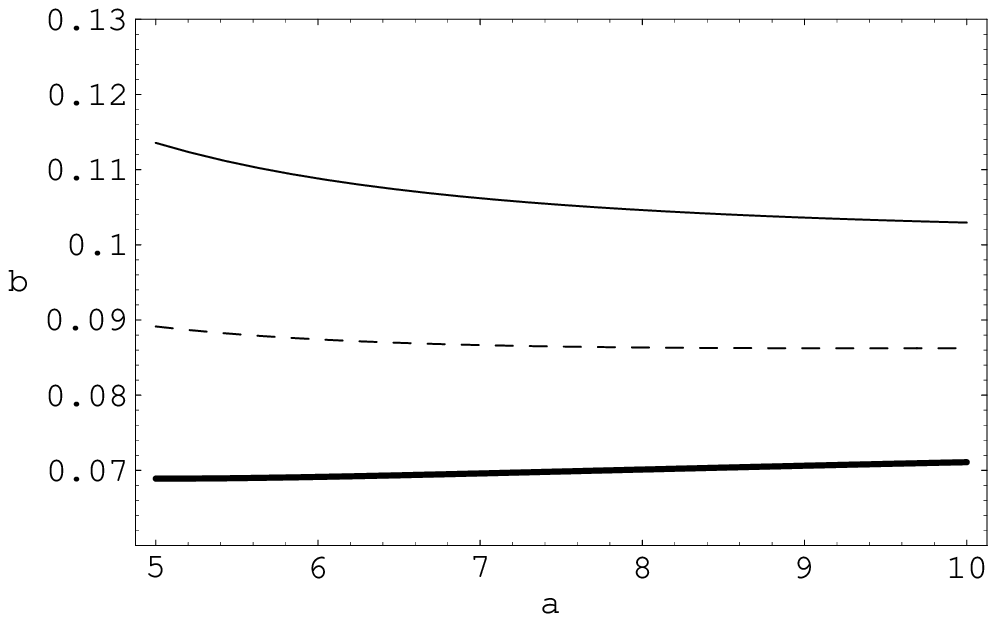}
\caption{\it{Stability plots for the sum rule in (\ref{eq:FSR}) of the Borel parameters for $B \to K_1(1400)\gamma$ at $s_0=33\ GeV^2$. Solid, Dashed and Thick lines correspond respectively to $m_b$=4.6 GeV, 4.7 GeV and 4.8 GeV.}}
\label{33F10ofK11400}
\end{center}
\end{figure}

\begin{figure}
%\psfrag{c}{\hskip 0.3cm $s(GeV^2)$}
\psfrag{b}{\hskip -2. cm $F_1^{K^*(1410)}(0)$}
\psfrag{a}{\hskip 0.3cm $t(GeV^2)$}
\begin{center}
\includegraphics[width=12cm,height=9cm]{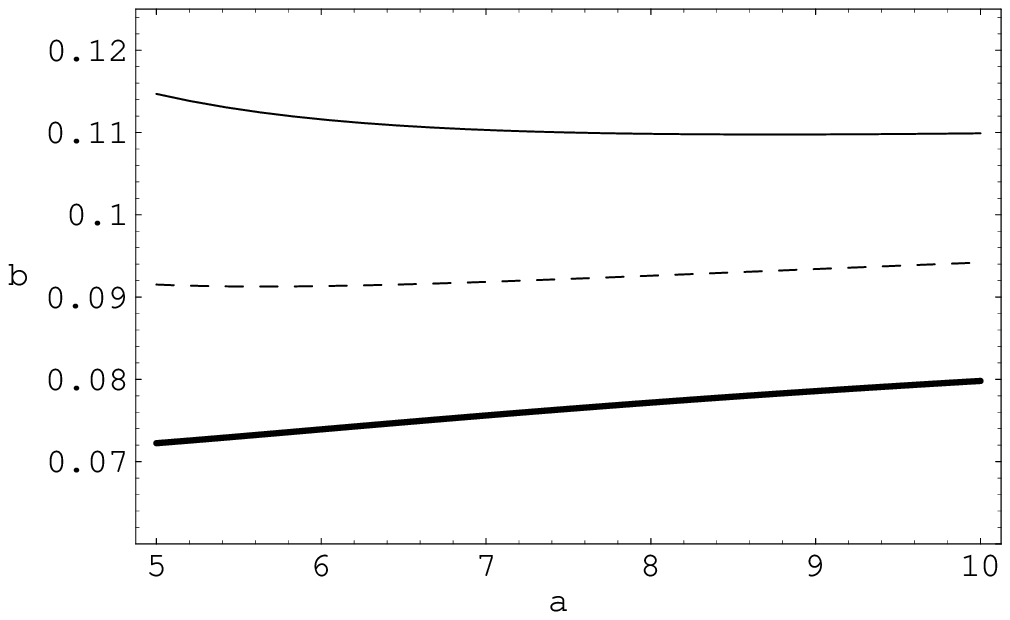}
\caption{\it{Stability plots for the sum rule in (\ref{eq:FSR}) of the Borel parameters for $B \to K^*(1410)\gamma$ at $s_0=35\ GeV^2$. Solid, Dashed and Thick lines correspond respectively to $m_b$=4.6 GeV, 4.7 GeV and 4.8 GeV.}}
\label{F10ofKstar1410}
\end{center}
%\end{figure}
\vspace{1.5cm}
%\begin{figure}
%\psfrag{c}{\hskip 0.3cm $s(GeV^2)$}
\psfrag{b}{\hskip -2. cm $F_1^{K^*(1410)}(0)$}
\psfrag{a}{\hskip 0.3cm $t(GeV^2)$}
\begin{center}
\includegraphics[width=12cm,height=9cm]{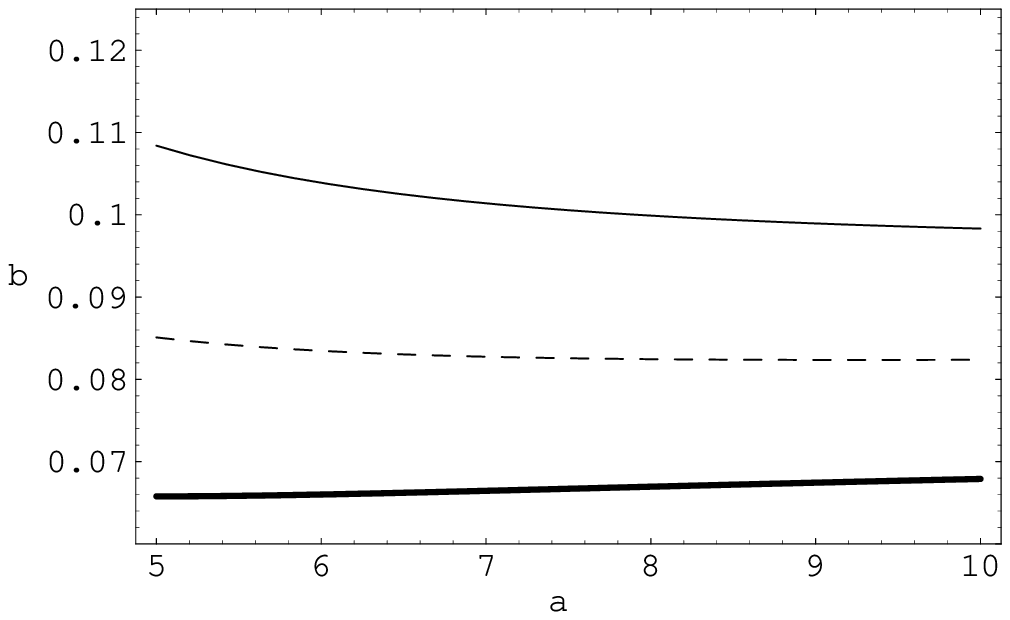}
\caption{\it{Stability plots for the sum rule in (\ref{eq:FSR}) of the Borel parameters for $B \to K^*(1410)\gamma$ at $s_0=33\ GeV^2$. Solid, Dashed and Thick lines correspond respectively to $m_b$=4.6 GeV, 4.7 GeV and 4.8 GeV.}}
\label{33F10ofKstar1410}
\end{center}
\end{figure}

\begin{figure}
%\psfrag{c}{\hskip 0.3cm $s(GeV^2)$}
\psfrag{b}{\hskip -2. cm $F_1^{K_1(1650)}(0)$}
\psfrag{a}{\hskip 0.3cm $t(GeV^2)$}
\begin{center}
\includegraphics[width=12cm,height=9cm]{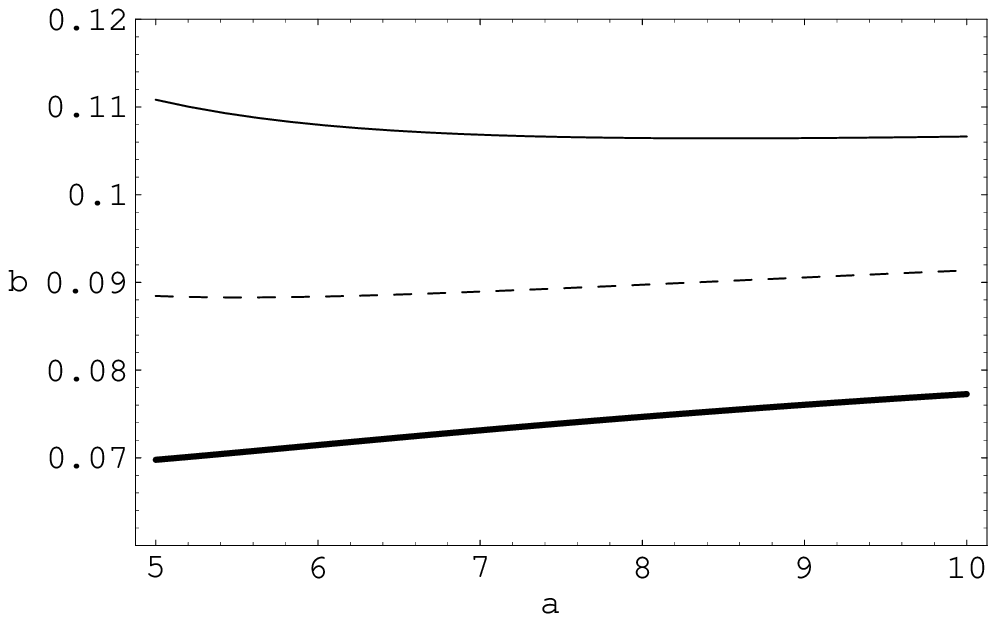}
\caption{\it{Stability plots for the sum rule in (\ref{eq:FSR}) of the Borel parameters for $B \to K_1(1650)\gamma$ at $s_0=35\ GeV^2$. Solid, Dashed and Thick lines correspond respectively to $m_b$=4.6 GeV, 4.7 GeV and 4.8 GeV.}}
\label{F10ofK11650}
\end{center}
%\end{figure}
\vspace{1.5cm}
%\begin{figure}
%\psfrag{c}{\hskip 0.3cm $s(GeV^2)$}
\psfrag{b}{\hskip -2. cm $F_1^{K_1(1650)}(0)$}
\psfrag{a}{\hskip 0.3cm $t(GeV^2)$}
\begin{center}
\includegraphics[width=12cm,height=9cm]{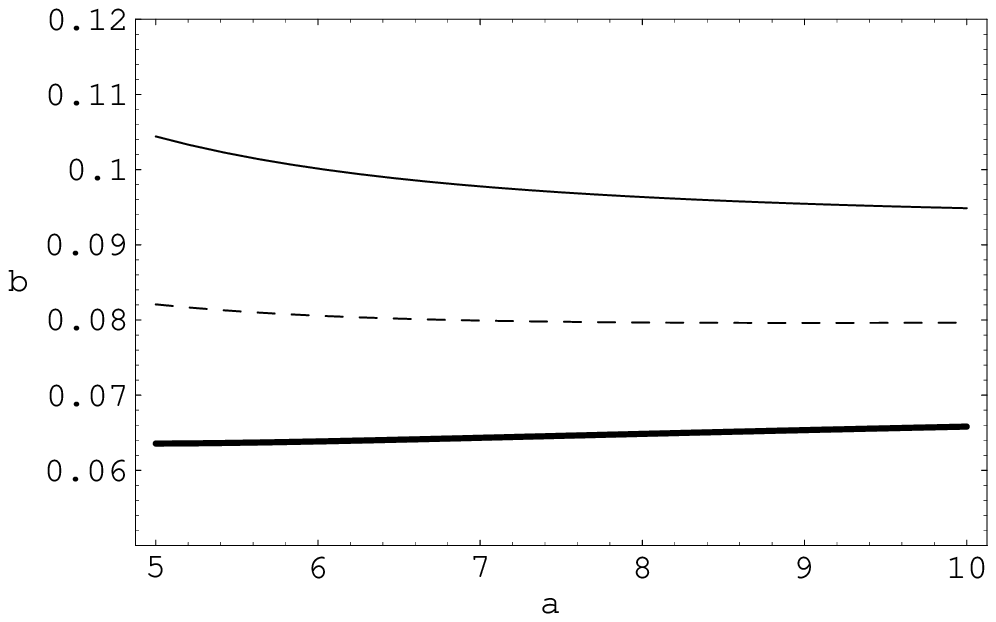}
\caption{\it{Stability plots for the sum rule in (\ref{eq:FSR}) of the Borel parameters for $B \to K_1(1650)\gamma$ at $s_0=33\ GeV^2$. Solid, Dashed and Thick lines correspond respectively to $m_b$=4.6 GeV, 4.7 GeV and 4.8 GeV.}}
\label{33F10ofK11650}
\end{center}
\end{figure}

\begin{figure}
%\psfrag{c}{\hskip 0.3cm $s(GeV^2)$}
\psfrag{b}{\hskip -2. cm $F_1^{K^*(1680)}(0)$}
\psfrag{a}{\hskip 0.3cm $t(GeV^2)$}
\begin{center}
\includegraphics[width=12cm,height=9cm]{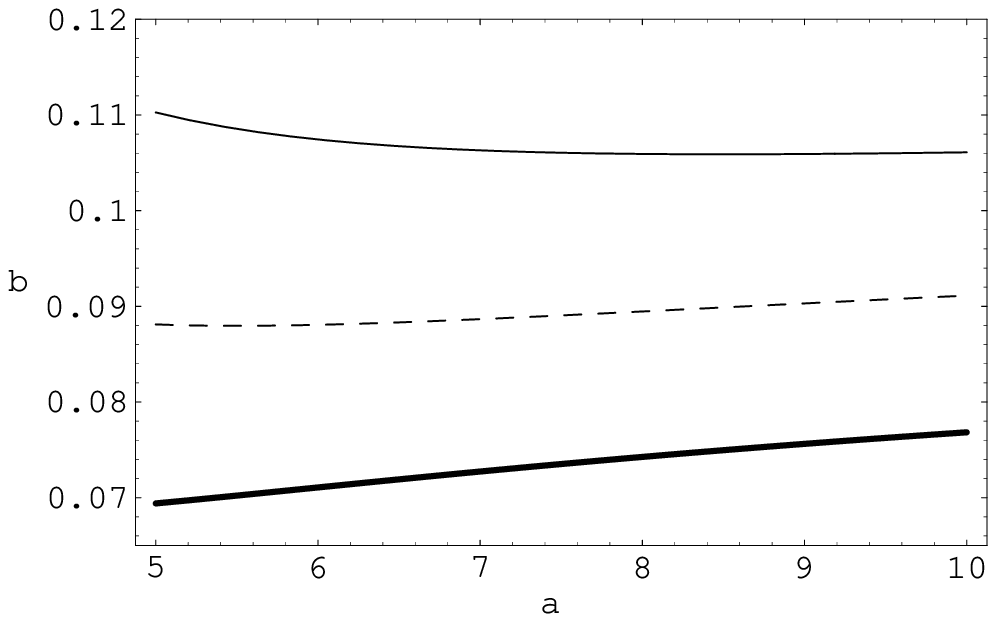}
\caption{\it{Stability plots for the sum rule in (\ref{eq:FSR}) of the Borel parameters for $B \to K^*(1680)\gamma$ at $s_0=35\ GeV^2$. Solid, Dashed and Thick lines correspond respectively to $m_b$=4.6 GeV, 4.7 GeV and 4.8 GeV.}}
\label{F10ofKstar1680}
\end{center}
%\end{figure}
\vspace{1.5cm}
%\begin{figure}
%\psfrag{c}{\hskip 0.3cm $s(GeV^2)$}
\psfrag{b}{\hskip -2. cm $F_1^{K^*(1680)}(0)$}
\psfrag{a}{\hskip 0.3cm $t(GeV^2)$}
\begin{center}
\includegraphics[width=12cm,height=9cm]{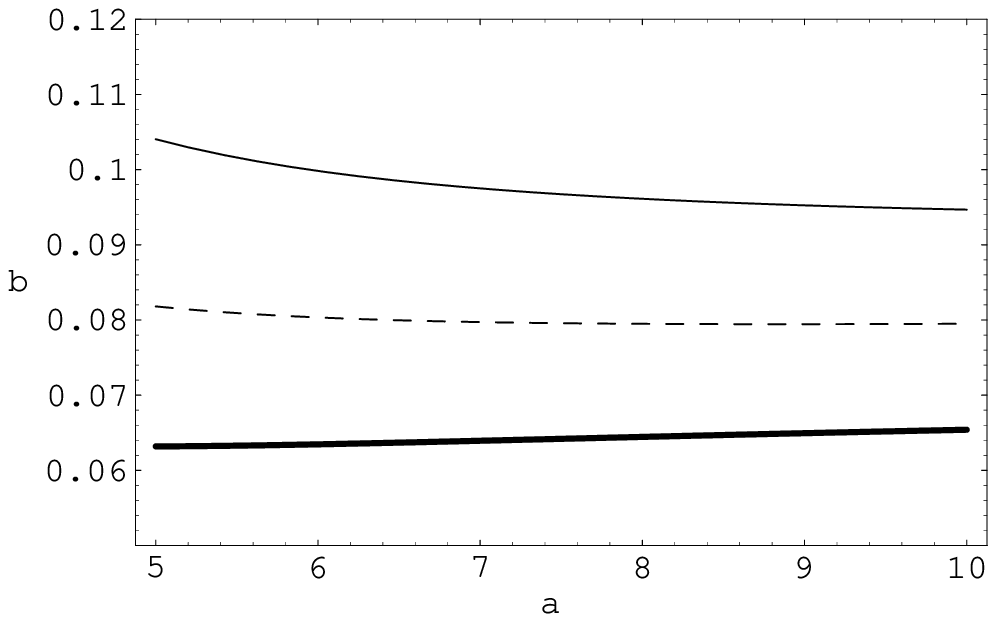}
\caption{\it{Stability plots for the sum rule in (\ref{eq:FSR}) of the Borel parameters for $B \to K^*(1680)\gamma$ at $s_0=33\ GeV^2$. Solid, Dashed and Thick lines correspond respectively to $m_b$=4.6 GeV, 4.7 GeV and 4.8 GeV.}}
\label{33F10ofKstar1680}
\end{center}
\end{figure}

\begin{figure}
\psfrag{c}{\hskip 0.3cm $s(GeV^2)$}
\psfrag{b}{\hskip -2. cm $F_1^{K^*_2(1430)}(0)$}
\psfrag{a}{\hskip 0.3cm $t(GeV^2)$}
\begin{center}
\includegraphics[width=12cm,height=9cm]{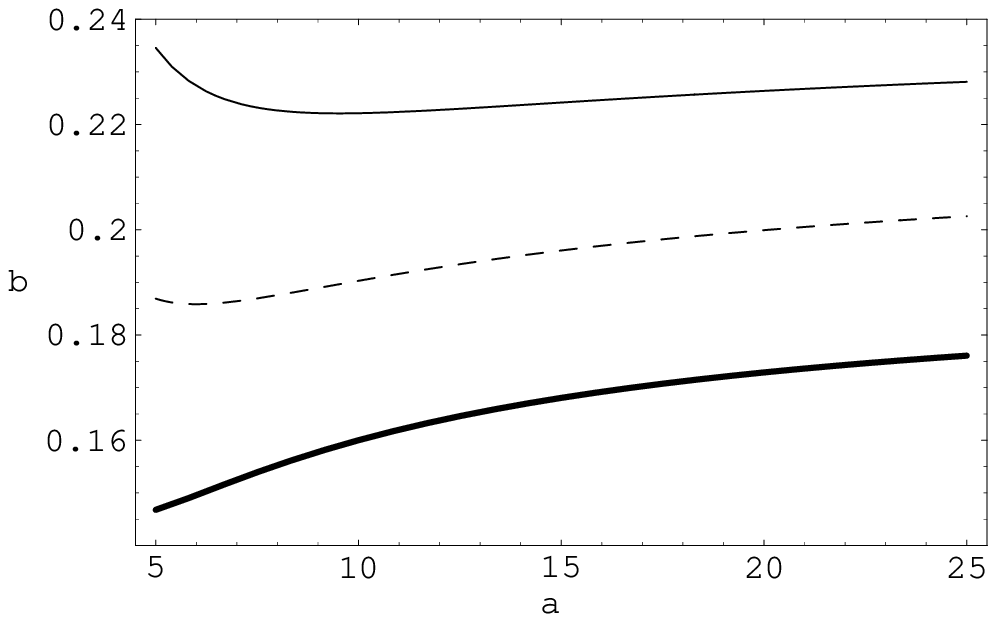}
\caption{\it{Stability plots for the sum rule in (\ref{eq7}) of the Borel parameters for $B \to K^*_2(1430)\gamma$ at $s_0=35\ GeV^2$. Solid, Dashed and Thick lines correspond respectively to $m_b$=4.6 GeV, 4.7 GeV and 4.8 GeV.}}
\label{35F10ofK21430}
\end{center}
%\end{figure}
\vspace{1.5cm}
%\begin{figure}
\psfrag{c}{\hskip 0.3cm $s(GeV^2)$}
\psfrag{b}{\hskip -2. cm $F_1^{K^*_2(1430)}(0)$}
\psfrag{a}{\hskip 0.3cm $t(GeV^2)$}
\begin{center}
\includegraphics[width=12cm,height=9cm]{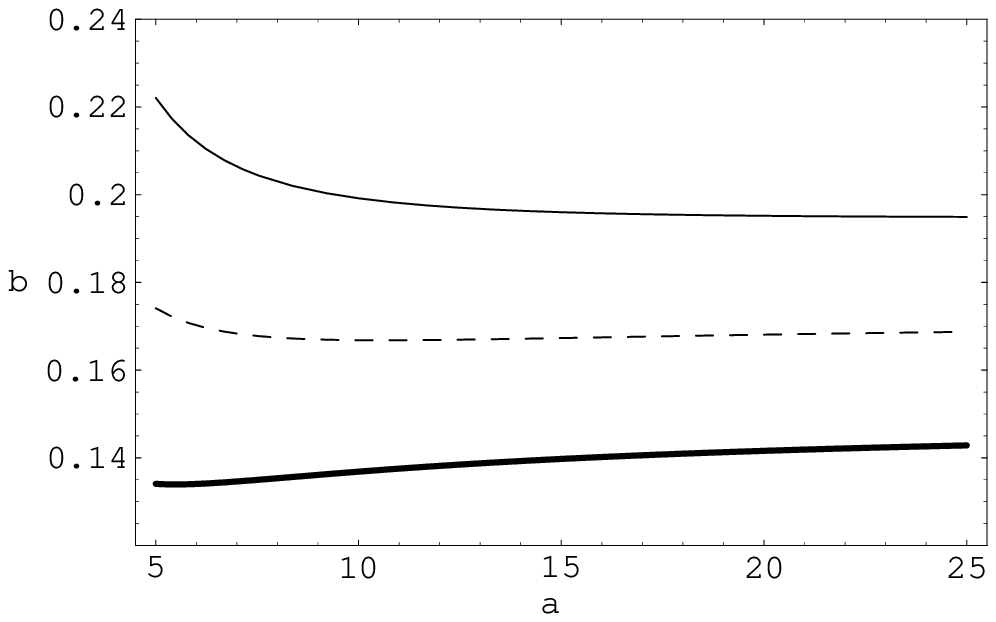}
\caption{\it{Stability plots for the sum rule in (\ref{eq7}) of the Borel parameters for $B \to K^*_2(1430)\gamma$ at $s_0=33\ GeV^2$. Solid, Dashed and Thick lines correspond respectively to $m_b$=4.6 GeV, 4.7 GeV and 4.8 GeV.}}
\label{33F10ofK21430}
\end{center}
\end{figure}

In Figs.(\ref{F10ofK21580}) and (\ref{25F10ofK21580}), we have plotted the normalized branching ratio of the $K_2(1580)$-meson as function of its decay constant. The shaded area depicts mainly the $m_b$-related uncertainty in our sum rule (\ref{eq7}) at $t=20\ GeV^2$ and $t=25\ GeV^2$. It turns out from Figs.(\ref{F10ofK21580}) and (\ref{25F10ofK21580}) that the uncertainty on the corresponding normalized branching ratio increases rapidly with the decay constant. 

The given errors in (\ref{F1V(A)}) and (\ref{F1T}) mainly come from the uncertainties in the $b$-quark mass and the dependence on the Borel parameter. This leads for all form factors except $F_1^{K_2(1580)}$, to an uncertainty of order $20\%$.  We observe that for all decays the average value of the momentum fraction $u$ under the integral in (\ref{eq:FSR}) and (\ref{eq7}) is $<u>\simeq 0.6-0.7$. It should be noted that errors given here reflect uncertainties in the input parameters, but do not include a possible theoretical uncertainty of the method itself.

Using our form factors (\ref{F1V(A)}) and (\ref{F1T}), we estimate  the  corresponding ratio $R= \Gamma(B\to K^{**}\gamma)/ \Gamma(B\to X_s\gamma)$ for various   $K^{**}$-mesons. We find a substantial fraction $(6.0-14)\%$ of the inclusive $b\to s\gamma$ branching ratio going into the $K^{*}(892)$ channel, and  $(3.0-7.0)\%$ going into the $K^*_2(1430)$ channel. Both of these predictions are in good agreement with the recent experimental result of $R_{K^{*}(892)}\approx(17.59-11.63)\%$ and $R_{K^*_2(1430)}\approx (4.23-2.98)\%$, presented in (\ref{Brk*0}), (\ref{BrXs}) and  (\ref{Brk2*}).

In order to make comparison of our results with previous calculations, we have tabulated our results together with results of \cite{altomari}, \cite{mannel}, \cite{Veseli}  and \cite{Faustov}  in table \ref{tab2}. As far as we know, these four works are the only ones that have dealt with radiative rare $B$ decays into higher $K$-resonances. Concerning $B\to K^{*}(892)\gamma$ and  $B\to K^*_2(1430)\gamma$, our predictions agree with the values quoted in \cite{Veseli} and \cite{Faustov}. Note that our result are in general in much better agreement with \cite{Veseli} than \cite{altomari} and \cite{mannel}.

With the exception of the $K^{*}(892)\gamma$ and $K^*_2(1430)\gamma$ channels, no other exclusive radiative processes have been identified so far. Thus, it is difficult to make more quantitative statements on the other $K^{**}$-decay channels, especially when the decay in consideration is totally dominated by the long distance effects.

\newpage \pagestyle{plain}

\begin{table}
\begin{center}
%\begin{tabular}{|c|}
%\hline  $R_{F}\times 10^2$ \\
\begin{tabular}{|cc|ccccc|}
\hline   & & & &$R_{F}[\%]$ & &\\
\hline
Meson   &   $J^{P}$   & (our)   &  ref.\cite{Veseli} &  ref.\cite{mannel} &  ref.\cite{altomari} &  ref.\cite{Faustov}   \\
\hline
$K$    & $0^{-}$  & \multicolumn{4}{c|}{forb.} & \\
$K^{*}(892)$ &  $1^{-}$ &$10.0\pm 4.0$ &
$16.8\pm 6.4$ & $3.5-12.2$ & 4.5 &$15\pm 3$\\
$K^{*}(1430)$ &  $0^{+}$& 
 \multicolumn{5}{c|}{forb.} \\
$K_{1}(1270)$ &  $1^{+}$ &$2.0\pm 0.8$ &
$4.3\pm 1.6$ & $4.5-10.1$ & forb./6.0 &$1.5\pm0.5$\\
$K_{1}(1400)$ &  $1^{+}$ &$0.9\pm 0.4$ &
$2.1\pm 0.9$ & $6.0-13.0$ &forb./6.0 &$2.6\pm0.6$\\
$K^{*}_{2}(1430)$ &  $2^{+}$ &$5.0\pm 2.0$ &
$6.2\pm 2.9$ & $17.3-37.1$ & 6.0&$5.7\pm 1.2$\\
$K^{*}(1680)$ &  $1^{-}$ &$0.7\pm 0.3$ &
$0.5\pm 0.2$ & $1.0-1.5$ & 0.9&\\
$K_{2}(1580)$ &  $2^{-}$ & &
$1.7\pm 0.4$ & $4.5-6.4$ & 4.4& \\
$K(1460)$    & $0^{-}$  & 
\multicolumn{4}{c|}{forb.}& \\
$K^{*}(1410)$ &  $1^{-}$ &$0.8\pm 0.4$ &
$4.1\pm 0.6$ & $7.2-10.6$ & 7.3&\\
$K^{*}_{0}(1950)$ &  $0^{+}$ & 
\multicolumn{4}{c|}{forb.}& \\
$K_{1}(1650)$ &  $1^{+}$ &$0.8\pm 0.3$ & $1.7\pm 0.6$ & not given& not given &\\
%\hline \multicolumn{3}{|c}{total} & $37.4\pm 13.6 $& $44.1 - 90.9$ & $29.1$ &
\hline
\end{tabular}
\end{center}
\caption{\it Comparison of our results for the ratio  $R_{F}[\%]$ with previous works.}\label{tab2}
\end{table}

%%%%%%%%%%%%%%%%%%%%%%%%%%%%%%%%%%%%%%%%%%%5
%\newpage \pagestyle{plain}
\section{Summary and Conclusions}
\hspace*{\parindent}
Motivated by the first observation of the radiative decay $B \to K^{*}_{2}(1430)\gamma$, we have investigated rare radiative $B$ decays to orbitally excited $K^{**}$-mesons. First, we have presented an alternative method of calculating the transition form factors and related decays using the QCD sum rules on the light-cone. We restricted our calculations to the leading twist-2 operator for the $K^*(892)$, as in \cite{AlietBraun}, and to the asymptotic wave function for the other $K^{**}$-mesons. The latter choice is simply based on the fact that using QCD sum rules, it is impossible to get rid of the lower-lying states contributions from these higher resonances.

However, nothing is known about the corresponding $K^{**}$-decay constants, and one has to predict them. For that, we have used recent data \cite{PDG} on semileptonic $\tau \to (P,S,V,A) \nu_{\tau}$ decays  to obtain them. 
For $K^{*}_{2}(1430)$, we have constrained the  corresponding decay constant with the recent data \cite{CLEO} on $B\to K^{*}_{2}(1430)\gamma$. We find that if $f_{K^{*}_{2}(1430)}= (140-180) \ MeV$, a substantial fraction $(3.0-7.0)\%$ of the inclusive $b\to s \gamma$ branching ratio goes into the $K^{*}_{2}(1430)$ channel, in a good agreement with recent CLEO data \cite{CLEO}. Our prediction for the $B \to K^{*}(892)\gamma$ branching fraction yields to $(6.0-14.0)\%$, also in good agreement with the experimental data \cite{CLEO, BABAR, BELLE}. 
As far as decays into higher $K$-resonances are concerned, our results are in general in much better agreement with \cite{Veseli} than \cite{altomari} and \cite{mannel}, apart from the $K^{*}(1410)$-channel where the difference is more significant. For the $K_{2}(1580)$-meson, we have plotted $Br(B\to K_{2}(1580)\gamma)/Br(B\to X_s\gamma)$ as function of the corresponding decay constant.

Finally, it should be noticed that the theoretical uncertainties in our  light-cone sum rules are the wave functions and the decay constants of the  $K^{**}$-mesons. The accuracy of our calculation can be substantially improved by taking into account the wave functions of twist-3 and twist-4 for the $B \to K^{*}(892)\gamma$ decay, and going beyond the asymptotic form for the other decay modes. To reduce the uncertainties on the $K^{**}$-mesons decay constants, one can determine them independently using QCD sum rules for the two-point correlator of the corresponding currents.

%\newpage \pagestyle{plain}
\vskip 5mm %------------------------------------------------------

\section*{Acknowledgments}

\hspace*{\parindent} I would like to thank the German Academic Exchange Service (DAAD) for financial support. I express my gratitude to A. Ali for suggesting this problem, for stimulating discussions and useful remarks on the manuscript. I am grateful to V.M. Braun 
 and A. Khodjamirian for several discussions.

%%%%%%%%%%%%%%%%%%%%%%%%%%%%%%%%%%%%%


\begin{thebibliography}{99}
\bibitem{CLEO} CLEO Collaboration, T. E. Coan et al.,Phys. Rev. Lett. 84 (2000) 5283.
\bibitem{BABAR} T. Pulliam, [BABAR Collaboration], talk presented at PHENO 2001, Madison, Wisconsin, May 2001, BaBar-Talk-01/53.
\bibitem{BELLE}G. Taylor, [BELLE Collaboration], talk presented at the $36^{th}$ Rencontres de Moriond, Electroweak Interactions and Unified Theories, Les Arcs, France, March 2001.
\bibitem{CLEO2}CLEO Collaboration, S. Ahmed et al., hep-ex/9908022; CLEO Collaboration, S. Chen et al., hep-ex/0108032.
\bibitem{ALEPH} ALEPH Collaboration, R. Barate et al., Phys. Lett. {\bf B429} (1998) 169.
\bibitem{BELLE2} BELLE Collaboration, Y. Ushiroda, hep-ex/0104045. 


%%%%%%%%%%%%%%%%%%%%%%%%%%%%%%%%%%%%%%%%%%%%%

\bibitem{ali2} A. Ali, Nucl. Instrum. Meth. {\bf A462} (2001) 11. 
\bibitem{altomari} T. Altomari,
Phys. Rev. D {\bf 37}, 677 (1988). 
\bibitem{mannel} A. Ali, T. Ohl, and T. Mannel,
Phys. Lett. {\bf B298}, 195 (1993).
\bibitem{Veseli}S. Veseli and M.G. Olsson, Phys.Lett.{\bf B367}(1996)309. 
\bibitem{Faustov} D. Ebert, R. N. Faustov, V.O. Galkin and H. Toki, 
Phys. Lett. {\bf B495} (2000) 309; Phys.Rev.{\bf D64} (2001) 054001. 

%%%%%%%%%QCD sum rules ref.%%%%%%%%%%%
\bibitem{Shifman} M. A. Shifman, A.I. Vainstein and V.I. Zakharov, Nucl. Phys.{\bf B147} (1979) 385. 
\bibitem{Balitsky} I. I. Balitsky,V. M. Braun and A. V. Kolesnichenko, Sov. J. Nucl. Phys. 44 (1986) 1028;   Nucl. Phys.{\bf B312} (1989) 509.
\bibitem{Braun}V. M. Braun and I. E. Filyanov, Z. Phys. {\bf C48} (1990) 239.
\bibitem{Chernyak} V. L. Chernyak, A. R. Zhitnisky, Phys. Rep. 112 (1984) 173.
\bibitem{Gorsky} A. S.Gorsky  Sov. J. Nucl. Phys. 41 (1985) 1008; ibid. 45 (1987) 512.
\bibitem{AlietGreub} A. Ali and C. Greub, Z. Phys. {\bf C49} (1991) 431.
\bibitem{Buras} A. J. Buras et al. Nucl. Phys. {\bf B424} (1994) 374.
\bibitem{munz}K. Chetyrkin, M. A. Misiak and M\"{u}nz,  Phys. Lett. {\bf B400} (1997) 206; Erratum-ibid {\bf B425} (1998) 414.
\bibitem{PDG} Particle Data Group, Lepton Summary Table, Eur. Phys. J. {\bf C15} (2000) 23. 
\bibitem{Khodjamirian} V. M. Belyaev, A. Khodjamirian and R. R\"{u}ckl, Z. Phys. {\bf C60} (1993) 349.
\bibitem{AlietBraun} A. Ali, V. M. Braun and H. Simma, Z. Phys. {\bf C67} (1994) 437.
\bibitem{Wandzura}S. Wandzura and F. Wilczek,  Phys. Lett. {\bf B82} (1977) 195.
\bibitem{Zhitnisky}A. R. Zhitnisky, A.I. Zhitnisky, and V. L. Chernyak, Sov. J. Nucl. Phys. 41 (1985) 284.

\bibitem{Reinders1}L. J. Reinders, H.R. Rubinstein, and S. Yazaki, Phys. Lett. {\bf B104} (1981) 305.
\bibitem{Aliev} T. M. Aliev and V.L. Eletsky, Sov. J. Nucl. Phys. {\bf 38} (1983) 936.
\bibitem{Narison}S. Narison,  Phys. Lett. {\bf B198} (1987) 104.

\bibitem{Reinders2}L. J. Reinders, Phys. Rev. {\bf D38} (1988) 947.
\bibitem{Dominguez}C. A. Dominguez and N. Paver Phys. Lett. {\bf B269} (1991) 169.
\bibitem{Bagan}E. Bagan, P.Ball, V.M. Braun and H.G. Dosch,Phys. Lett. {\bf B278} (1992) 457.
\bibitem{Neubert} M. Neubert, Phys. Rev. {\bf D45} (1992) 2451.
\bibitem{Ruckl}A. Khodjamirian and R. R\"{u}ckl, {\it{QCD sum rules for exclusive decays of heavy mesons}}, in Heavy Flavours II, eds. A. J. Buras and M. Linder, World Scientific, 1998.
\bibitem{Narison2}S. Narison,  Nucl. Phys. Proc. Suppl. {\bf 74} (1999) 304.

\bibitem{Jamin}M. Jamin and Bj\"{o}rn O. Lange, Preprint hep-ph/0108135.
\bibitem{Steinhauser} A. A. Penin and M. Steinhauser, Preprint hep-ph/0108110.

\bibitem{Donogue} J. F. Donogue, E. Golowich and B. R. Holstein, ``Dynamics of the standard model''(Cambridge university press, 1992).



\end{thebibliography}
\end{document}